\documentclass[10pt,final,journal,letterpaper,oneside,twocolumn]{IEEEtran}
\usepackage{physics}
\usepackage{tikz}
\usetikzlibrary{shapes.geometric, arrows.meta, positioning}

\tikzstyle{block} = [rectangle, draw, align=center, text centered,
                     minimum width=3.5cm, minimum height=1cm, fill=blue!10]
\tikzstyle{arrow} = [thick,->,>=stealth]
\usepackage{amsmath}
\usepackage{amsfonts}
\usepackage{mathtools}
\usepackage{amsfonts}
\usepackage{amssymb}
\usepackage{array}
\usepackage{makeidx}
\usepackage{csquotes}
\usepackage{graphicx}
\usepackage{cite}
\usepackage{xcolor}
\usepackage{hyperref}
\usepackage{subcaption}
\usepackage{graphicx}
\usepackage{multicol}
\usepackage{multirow}
\usepackage{pifont}

\usepackage[table]{xcolor}
\usepackage{array}
\usepackage{tabularx}
\usepackage{booktabs}

\definecolor{HeaderBlue}{RGB}{195,225,230}
\definecolor{BodyBlue}{RGB}{235,247,249}

\definecolor{UElight}{RGB}{240, 245, 255} 
\definecolor{UEmedlight}{RGB}{210, 225, 250} 
\definecolor{UEmedium}{RGB}{180, 200, 240}
\definecolor{UEmeddark}{RGB}{140, 170, 230}
\definecolor{UEdark}{RGB}{100, 140, 210}   
\definecolor{CodingRed}{RGB}{245, 210, 210} 

\newcolumntype{L}{>{\raggedright\arraybackslash}X}

\usepackage{tikz}
\usetikzlibrary{arrows.meta, positioning}

\definecolor{HeaderRed}{RGB}{245,210,210}   
\definecolor{SubGreen}{RGB}{232,245,232}    
\definecolor{BusLine}{RGB}{120,120,120}     

\usetikzlibrary{shapes.symbols, positioning, arrows.meta, calc}

\usetikzlibrary{shapes.geometric, arrows.meta, positioning, calc}

\definecolor{signalblue}{RGB}{0,153,255}
\definecolor{signalred}{RGB}{200,30,30}
\definecolor{signalorange}{RGB}{255,140,0}
\definecolor{carbody}{RGB}{255,200,50}
\definecolor{towergray}{RGB}{80,80,80}

\usetikzlibrary{backgrounds}

\usepackage[utf8]{inputenc}
\usepackage{tikz}
\usepackage{tikz-3dplot}
\usetikzlibrary{backgrounds,fit}

\definecolor{BodyBlue}{RGB}{235,247,249}

\begin{document}
 
\title{\LARGE A Survey on STAR-RIS Enabled Joint Communications and Sensing: Fundamentals, Recent Advances and Research Challenges}

\author{Wali Ullah Khan,~\IEEEmembership{Member,~IEEE} Chandan Kumar Sheemar,~\IEEEmembership{Member,~IEEE} Syed Tariq Shah,~\IEEEmembership{Member,~IEEE} \\ Manzoor Ahmed,~\IEEEmembership{Member,~IEEE}, and Symeon Chatzinotas,~\IEEEmembership{Fellow,~IEEE}  \thanks{Wali Ullah Khan, Chandan Kumar Sheemar, and Symeon Chatzinotas are with Interdisciplinary Centre for Security, Reliability and Trust (SnT), University of Luxembourg, Luxembourg
(emails: \{waliullah.khan, chandankumar.sheemar, jorge.querol, symeon.chatzinotas\}@uni.lu).

Syed Tariq Shah is with the School of Computer Science and Electronic Engineering, University of Essex, Colchester, UK (email: syed.shah@essex.ac.uk).

Manzoor Ahmed is with the School of Computer and Information Science and also with Institute for AI Industrial Technology Research, Hubei Engineering University, Xiaogan City, 432000, China  (email: manzoor.achakzai@gmail.com).

}}%

\markboth{IEEE Internet of Things Journal (For Review)
}
{Shell \MakeLowercase{\textit{et al.}}: Bare Demo of IEEEtran.cls for IEEE Journals} 

\maketitle

\begin{abstract}
The joint communications and sensing (JCAS) paradigm is envisioned as a core capability of sixth-generation (6G) wireless networks, enabling the integration of data communication and environmental sensing within a unified system. By reusing spectrum, waveforms, and hardware resources, JCAS improves spectral efficiency, reduces system complexity, and hardware cost, while enabling new use cases. Nevertheless, the realization of JCAS is hindered by inherent trade-offs between communication and sensing objectives, limited controllability of wireless propagation, and stringent hardware and design constraints. Simultaneously transmitting and reflecting reconfigurable intelligent surfaces (STAR-RIS) have recently emerged as a promising technology to address these challenges by enabling full-space programmable manipulation of electromagnetic waves. This survey provides a systematic and in-depth review of STAR-RIS-enabled JCAS systems. Specifically, we first introduce the fundamental principles of JCAS and STAR-RIS. We then classify and review the state-of-the-art research on STAR-RIS-assisted JCAS from multiple perspectives, encompassing system architectures, waveform and beamforming design, resource allocation, optimization frameworks, and learning-based control. Finally, we identify key open challenges that remain unsolved and outline promising future research directions toward intelligent, flexible, and perceptive 6G wireless networks.

\end{abstract}


\IEEEpeerreviewmaketitle

\section{Introduction}
\IEEEPARstart{T}{he} rapid evolution of wireless networks towards sixth-generation (6G) systems is driven by the growing demand for ubiquitous connectivity, ultra-high data rates, ultra-low latency, and native intelligence \cite{dang2020should,tariq2020speculative,sheemar2021hybrid}. Beyond traditional communication-centric objectives, future wireless networks are envisioned to provide an integrated platform capable of perceiving, interpreting, and interacting with the physical environment. This paradigm shift has brought joint communications and sensing (JCAS) \cite{fang2022joint,sheemar2025joint}, also called integrated sensing and communication (ISAC) \cite{lu2024integrated}, to the forefront of 6G research. By enabling simultaneous data transmission and environmental sensing using shared wireless resources, JCAS promises significant improvements in spectral efficiency, hardware utilization, and system intelligence \cite{sheemar2023full}.

JCAS is motivated by emerging applications such as autonomous driving, smart cities, industrial automation, extended reality (XR), and unmanned aerial systems \cite{ma2023integrated,bovzanic20216g}, where reliable communication and real-time situational awareness are equally critical. Conventionally, communication and sensing systems have been designed and deployed independently, leading to redundant hardware, inefficient spectrum usage, and increased operational complexity. In contrast, JCAS leverages the dual-use nature of radio-frequency (RF) signals, allowing the same waveform, spectrum, and transceiver infrastructure to support both functions \cite{liu2020joint,sheemar2025joint,sheemar2025joint_mag}. As a result, JCAS is widely regarded as a foundational capability for future 6G networks \cite{parssinen2021white}. Despite its promising potential, the practical realization of JCAS faces several fundamental challenges \cite{sheemar2023fullFD}. These include conflicting performance requirements between communication and sensing, complex multi-objective waveform and beamforming design, stringent hardware constraints, and sensitivity to environmental dynamics such as blockage and interference \cite{liu2020joint_sur}. Addressing these challenges requires new degrees of freedom in wireless system design that go beyond conventional active transceiver architectures.

Reconfigurable Intelligent Surfaces (RIS) are implemented using metasurfaces, which have evolved from passive reflectors to programmable surfaces capable of dynamically altering electromagnetic wave propagation \cite{li2025ris,khan2024reconfigurable,iacovelli2024holographic,sheemar2025UAV,khan2024beyondIoT}. Traditional RIS focuses mainly on reflection, optimizing wireless signals by adjusting phase shifts \cite{khan2024beyond}. Simultaneously Transmitting and Reflecting RIS (STAR-RIS) extends this concept by enabling transmission and reflection, allowing full-space coverage and enhanced wireless performance \cite{mu2021simultaneously}. STAR-RIS represents an advanced version of classical reflective RIS by supporting dual functionalities and improving energy efficiency, coverage, and adaptability in next-generation networks \cite{wu2021coverage}.

\begin{table*}[t]\scriptsize
\caption{Complete List of Abbreviations}
\label{tab:abbreviations}
\centering
\setlength{\tabcolsep}{5pt}
\renewcommand{\arraystretch}{1.1}

\definecolor{HeaderRed}{RGB}{245,210,210}


\begin{tabularx}{\textwidth}{|
p{1cm} L
!{\vrule width 0.6pt}
p{1cm} L
!{\vrule width 0.6pt}
p{1cm} L|}
\hline

\rowcolor{HeaderRed}
\textbf{Abbr.} & \textbf{Description} &
\textbf{Abbr.} & \textbf{Description} &
\textbf{Abbr.} & \textbf{Description} \\
\hline

\rowcolor{BodyBlue}
6G & Sixth-generation wireless &
5G-A & 5G-Advanced &
AI & Artificial intelligence \\

\rowcolor{BodyBlue}
AirComp & Over-the-air computing &
AoA & Angle of arrival &
AoD & Angle of departure \\

\rowcolor{BodyBlue}
AN & Artificial noise &
AO & Alternating optimization &
AWGN & Additive white Gaussian noise \\

\rowcolor{BodyBlue}
BCD & Block coordinate descent &
B\&B & Branch-and-bound &
BS & Base station \\

\rowcolor{BodyBlue}
CE & Channel estimation &
CN & Complex normal distribution &
MUI & Multi-user interference \\

\rowcolor{BodyBlue}
CRB & Cram\'er--Rao bound &
CRLB & Cram\'er--Rao lower bound &
CSI & Channel state information \\

\rowcolor{BodyBlue}
CSF & Channel state feedback &
CU & Communication user &
D2D & Device-to-device \\

\rowcolor{BodyBlue}
DDPG & Deep deterministic policy gradient &
DFRC & Dual-functional radar--communication &
DoF & Degree(s) of freedom \\

\rowcolor{BodyBlue}
DoA & Direction of arrival &
DRL & Deep reinforcement learning &
EE & Energy efficiency \\

\rowcolor{BodyBlue}
EH & Energy harvesting &
ES & Energy splitting  &
EV & Eavesdropper \\

\rowcolor{BodyBlue}
FMCW & Frequency-modulated continuous-wave &
FP & Fractional programming &
ISAC & Integrated sensing and communication \\

\rowcolor{BodyBlue}
IoE & Internet of everything &
IoT & Internet of Things &
OMA & Orthogonal multiple access  \\

\rowcolor{BodyBlue}
JCAS & Joint communications and sensing &
JCAS & Joint sensing and communication  &
LoS & Line-of-sight \\

\rowcolor{BodyBlue}
MDP & Markov decision process &
MU & Multi-user &
MIMO & Multiple-input multiple-output \\

\rowcolor{BodyBlue}
MISO & Multiple-input single-output &
mmWave & Millimeter wave &
MS & Mode switching  \\


\rowcolor{BodyBlue}
NF & Near-field &
NMSE & Normalized mean squared error &
NTN & Non-terrestrial network \\

\rowcolor{BodyBlue}
OFDM & Orthogonal frequency-division multiplexing &
ISCC & Integrated sensing, communication, and control&
QCQP & Quadratically constrained quadratic program \\

\rowcolor{BodyBlue}
PAPR & Peak-to-average power ratio &
PDE & Probability of detection error &
NOMA & Non-orthogonal multiple access \\

\rowcolor{BodyBlue}
PZF & Partial zero-forcing &
OTFS & Orthogonal time frequency space &
QoS & Quality of service \\

\rowcolor{BodyBlue}
RCS & Radar cross section &
RF & Radio frequency &
RIS & Reconfigurable intelligent surface \\

\rowcolor{BodyBlue}
RL & Reinforcement learning &
RSMA & Rate-splitting multiple access &
SAC & Soft actor--critic \\

\rowcolor{BodyBlue}
SCMA & Sparse code multiple access &
SCNR & Signal-to-clutter-plus-noise ratio &
SDP & Semidefinite programming \\

\rowcolor{BodyBlue}
SDR & Semidefinite relaxation &
SE & Spectral efficiency &
SIC & Successive interference cancellation \\

\rowcolor{BodyBlue}
SINR & Signal-to-interference-plus-noise ratio &
SNR & Signal-to-noise ratio &
SPEB & Squared position error bound \\

\rowcolor{BodyBlue}
SCA & Successive convex approximation &
STAR & Simultaneously transmitting and reflecting &
SU & Sensing user / sensing target  \\


\rowcolor{BodyBlue}
STARS & Simultaneously transmitting and reflecting surface &
SWIPT & Simultaneous wireless information and power transfer &
TD3 & Twin delayed deep deterministic policy gradient \\

\rowcolor{BodyBlue}
THz & Terahertz &
T\&R & Transmission-and-reflection &
TS & Time switching (STAR-RIS protocol) \\

\rowcolor{BodyBlue}
UAV & Uncrewed aerial vehicle &
UE & User equipment &
URLLC & Ultra-reliable low-latency commun. \\

\rowcolor{BodyBlue}
V2I & Vehicle-to-infrastructure &
V2V & Vehicle-to-vehicle &
V2X & Vehicle-to-everything \\

\rowcolor{BodyBlue}
VEC & Vehicular edge computing &
VLC & Visible light communication &
WMMSE & Weighted minimum mean squared error \\

\rowcolor{BodyBlue}
WiGig & Wireless gigabit (IEEE 802.11ad/ay) &
XL & Extremely large (array/surface) &
ZF & Zero-forcing \\

\hline
\end{tabularx}
\end{table*}

\begin{table*}[t]
\caption{Related survey papers on JCAS-enabled 6G networks, where $\times$ means not covered, * means preliminary covered, and *** means fully covered.}
\label{tablecomp}
\centering
\setlength{\tabcolsep}{4pt}
\renewcommand{\arraystretch}{1.08}

\begin{tabular}{|p{0.55cm}|c|p{11cm}|c|c|c|}
\hline
\rowcolor{HeaderRed}
\textbf{Ref.} & \textbf{Year} & \textbf{Focus of the Related Surveys} & \textbf{STAR-RIS} & \textbf{JCAS} & \textbf{JCAS with STAR-RIS} \\ \hline
\rowcolor{BodyBlue}
\cite{9606831} & 2021 & \raggedright Highlights JCAS's role in enabling IoT applications & $\times$ & *** & $\times$ \\ \hline
\rowcolor{BodyBlue}
\cite{9705498} & 2022 & \raggedright Explores theoretical trade-offs and performance limits of JCAS systems & $\times$ & *** & $\times$ \\ \hline
\rowcolor{BodyBlue}
\cite{9924202} & 2022 & \raggedright Explores joint waveform design to balance JCAS sensing and communication & $\times$ & *** & $\times$ \\ \hline
\rowcolor{BodyBlue}
\cite{9919739} & 2022 & \raggedright Discusses JCAS's potential in advanced vehicular networks & $\times$ & *** & $\times$ \\ \hline
\rowcolor{BodyBlue}
\cite{10944644} & 2023 & \raggedright JCAS as a solution to CAV network challenges for 6G & $\times$ & *** & $\times$ \\ \hline
\rowcolor{BodyBlue}
\cite{10098686} & 2023 & \raggedright UAVs' role in JCAS for real-time adaptability and sensing flexibility & $\times$ & *** & $\times$ \\ \hline
\rowcolor{BodyBlue}
\cite{10012421} & 2023 & \raggedright Signal design and optimization for JCAS in 5G-A and 6G & $\times$ & *** & $\times$ \\ \hline
\rowcolor{BodyBlue}
\cite{salem2023data} & 2023 & \raggedright Data-driven approaches to enhance JCAS efficiency and accuracy & $\times$ & *** & $\times$ \\ \hline
\rowcolor{BodyBlue}
\cite{ahmed2023survey} & 2023 & \raggedright Comprehensive survey on STAR-RIS advances & *** & $\times$ & $\times$ \\ \hline
\rowcolor{BodyBlue}
\cite{liu2023simultaneously} & 2023 & \raggedright Survey on STAR-RIS enabled 6G networks & *** & $\times$ & $\times$ \\ \hline
\rowcolor{BodyBlue}
\cite{10756650} & 2024 & \raggedright Role of ML in improving sensing, communication, and resource allocation & $\times$ & *** & $\times$ \\ \hline
\rowcolor{BodyBlue}
\cite{10663823} & 2024 & \raggedright Advances in JCAS through sensing, communication, and AI collaboration & $\times$ & *** & $\times$ \\ \hline
\rowcolor{BodyBlue}
\cite{10489999} & 2024 & \raggedright Channel modeling approaches and opportunities for JCAS in 6G & $\times$ & *** & $\times$ \\ \hline
\rowcolor{BodyBlue}
\cite{10536135} & 2024 & \raggedright Vision of JCAS using signal processing, optimization, and ML for 6G & $\times$ & *** & $\times$ \\ \hline
\rowcolor{BodyBlue}
\cite{magbool2024survey} & 2024 & \raggedright Role of metasurfaces in enhancing JCAS adaptability & $\times$ & *** & $\times$ \\ \hline
\rowcolor{BodyBlue}
\cite{10812728} & 2024 & \raggedright Integration of sensing, communication, and computation in 6G & $\times$ & *** & $\times$ \\ \hline
\rowcolor{BodyBlue}
\cite{10770127} & 2024 & \raggedright Interference mitigation strategies in JCAS systems & $\times$ & *** & $\times$ \\ \hline
\rowcolor{BodyBlue}
\cite{shtaiwi2024orthogonal} & 2024 & \raggedright OTFS modulation's role in improving JCAS SE & $\times$ & *** & $\times$ \\ \hline
\rowcolor{BodyBlue}
\cite{10574259} & 2024 & \raggedright Enhancing security and privacy in JCAS systems & $\times$ & *** & $\times$ \\ \hline
\rowcolor{BodyBlue}
\cite{10494372} & 2024 & \raggedright Potential and challenges of using THz frequencies for JCAS & $\times$ & *** & $\times$ \\ \hline
\rowcolor{BodyBlue}
\cite{10735249} & 2024 & \raggedright Study the integration of JCAS with STAR-RIS & * & * & * \\ \hline
\rowcolor{BodyBlue}
\cite{iqbal2025comprehensive} & 2025 & \raggedright Presents a comprehensive survey of RIS and STAR-RIS technologies & *** & $\times$ & $\times$ \\ \hline

\rowcolor{BodyBlue}
\textbf{Our} & \textbf{2026} & \raggedright This paper provides a comprehensive survey on STAR-RIS enabled JCAS networks, covering fundamentals, architectural design and optimization strategies, waveform design, high mobility scenarios and UAV communications, security and privacy, NOMA/RSMA networks, and energy efficiency. It compares methods, discusses lessons, and identifies research directions. & *** & *** & *** \\ \hline
\end{tabular}
\end{table*}

From a fundamental perspective, STAR-RIS operation relies on carefully engineered meta-atoms that split the incident electromagnetic energy into transmitted and reflected components with controllable phase and amplitude responses \cite{zhao2022simultaneously}. Depending on the hardware implementation, this functionality can be realized through different operating modes, such as energy splitting, time switching, or mode switching \cite{ata2025time}. These modes introduce additional degrees of freedom for jointly optimizing communication and sensing performance, but they also impose practical constraints related to power conservation, hardware complexity, and control signaling. In terms of hardware architecture, STAR-RIS are typically composed of layered structures that enable independent control of transmission and reflection coefficients. Compared to conventional RIS, STAR-RIS generally require more sophisticated circuit designs and control mechanisms to achieve accurate and stable full-space manipulation \cite{mu2021simultaneously}. Nevertheless, they retain the key advantages of RIS technology, including low power consumption, lightweight structure, and ease of deployment on building facades, indoor walls, or mobile platforms. Compared to the conventional RIS, STAR-RIS is a much more promising candidate for enabling full space JCAS.

\subsection{Related Surveys and Main Contributions} Existing survey paper have explored JCAS from multiple perspectives, including enabling technologies, theoretical limits, design trade-offs, and application scenarios. In particular, the integration of JCAS within the Internet of Things (IoT) ecosystem has been highlighted as a means to enhance connectivity and spectrum efficiency \cite{9606831}, while other works have examined the fundamental trade-offs between sensing accuracy and communication reliability and proposed optimization frameworks to balance these objectives \cite{9705498}. A survey focused on waveform-level integration, commonly classifying JCAS designs into communication-centric, sensing-centric, and jointly optimized frameworks, with useful guidelines for system design, is available in \cite{9924202}. Application-driven studies have further examined the role of JCAS in vehicular networks and autonomous transportation, emphasizing ultra-low latency, high-accuracy localization, and safety-critical communication \cite{9919739,10944644}, as well as in UAV-assisted systems, where mobility and adaptability are key advantages \cite{10098686}.

With the evolution toward 6G, recent surveys have increasingly addressed signal processing, waveform design, and energy-efficient transmission for JCAS \cite{10012421}. The incorporation of machine learning and artificial intelligence has also been emphasized as a powerful tool for adaptive beamforming, localization, sensing enhancement, and resource allocation \cite{salem2023data}, and detailed comprehensive surveys are available in \cite{ahmed2023survey,10756650,liu2023simultaneously,10663823}. Additional works have considered practical challenges such as interference mitigation, spectrum sharing, and environmental dynamics \cite{10489999,10770127}, as well as emerging solutions including OTFS modulation for high-mobility scenarios \cite{shtaiwi2024orthogonal} and physical-layer security mechanisms \cite{10574259}. The potential of terahertz (THz) bands for JCAS has also been surveyed, highlighting both opportunities and implementation challenges \cite{10494372}. A magazine paper providing the fundamentals on STAR-RIS-enabled JCAS are discussed in the magazine paper \cite{10735249}. Finally, in \cite{iqbal2025comprehensive}, the authors partially cover the topic of STAR-RIS with particular focus on NOMA, but the potential of JCAS enabled by STAR-RIS is not discussed.

\begin{figure*}[!t]
\centering
\includegraphics[width=0.8\textwidth]{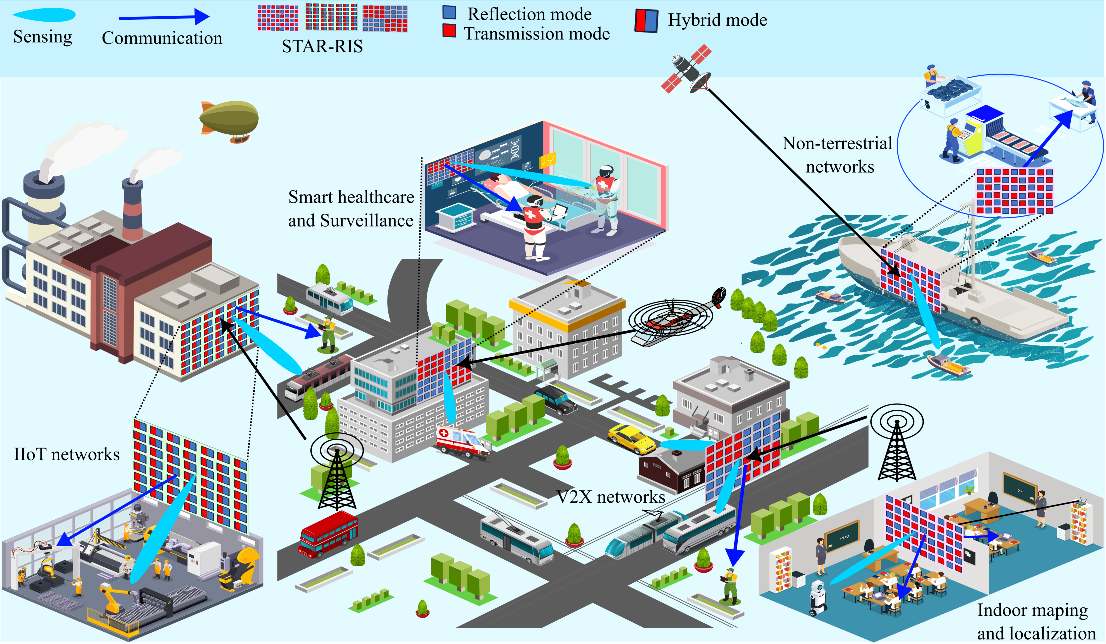}
\caption{Applications/Use cases of STAR-RIS enabled JCAS in 6G wireless networks.}
\label{STARRISApplication}
\end{figure*}

Table~\ref{tablecomp} summarizes representative survey works that investigate the role of JCAS with their main contributions to the state of the art. Overall, the existing survey literature provides a solid foundation for understanding JCAS systems and their evolution. However, a comprehensive surveys that explicitly focus on STAR-RIS-enabled JCAS, including full-space propagation control and joint sensing–communication design, is not available, which motivates the scope and contributions of this work.
Building on this observation, the objective of this survey is to provide a comprehensive, unified, and in-depth treatment of STAR-RIS-enabled JCAS networks. A general JCAS network assisted with STAR-RIS is shown in Figure \ref{STARRISApplication}.

To begin, this survey lays out the building blocks of signal models, operational procedures, hardware designs, and performance metrics as they pertain to dual function systems and STAR-RIS and JCAS. This specific survey focuses on properties unique to STAR-RIS; such as simultaneous transmission and reflection, energy splitting, and adjustable flexible mode control. These properties create more degrees of freedom for the end-user by offering more customizable trade-offs between the communication throughput and sensing accuracy. This survey moves on to provide a comprehensive, multi-faceted classiffication and review of the state of the art research on STAR-RIS and JCAS from a systems and network design; waveform and beamforming, resource distribution and optimization, and harnessing artificial intelligence for learning resource optimization in variable control environments. STAR-RIS is allied with JCAS to provide the most all encompassing design performance for a given capability and design context; this is particularly true for vehicular systems, UAV systems, indoor localization, and non-terrestrial network systems.

\begin{figure*}[!t]
\centering
\includegraphics[width=0.6\textwidth,height=8cm]{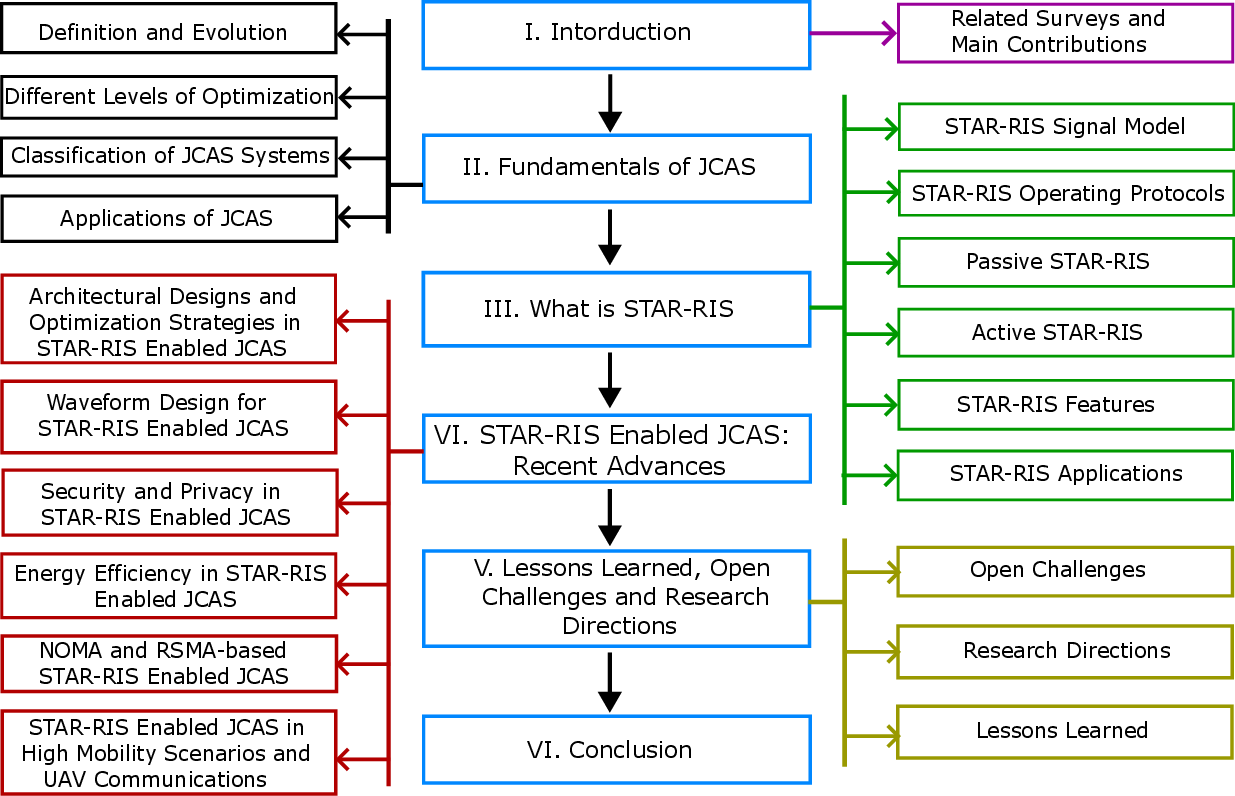}
\caption{Structure of this survey.}
\label{STARjcasTx}
\end{figure*}

Additionally, This survey outlines and examines specific technical issues that remain unsolved such as channel modeling and analysis for STAR-RIS-assisted JCAS, hardware limitations, scalability and control overhead, real-time optimization, and issues pertaining to security and privacy. This work attempts to provide a clear direction for future research by synthesizing existing outcomes and emphasizing current gaps. Overall, this survey aims to provide an all-encompassing source for JCAS, STAR-RIS, and 6G wireless system researchers and practitioners, and to assist in the creation of the intelligent, adaptive, and context-sensitive wireless networks made possible by STAR-RIS-enabled full-space JCAS systems. \emph{Paper Organization:} The overall structure of the survey is shown in Figure \ref{STARjcasTx}.

\section{Fundamentals of JCAS}
The systems of communication and systems of sensing have been evolving for decades in independent, parallel paths. Communication systems have been focused on the development of new techniques for transmitting data over varied distances and data channels. \cite{sheemar2022practical,sheemar2021game} Meanwhile, systems of sensing, including radar, lidar, and sonar, have been developed for the accurate sensing of the state of the environment, the detection of the objects, and the location of the objects. \cite{liang2011design} Each of the domains has been characterized by the development of highly specialized platforms, both in terms of hardware and in terms of techniques of signal processing, optimized for the goals of their particular domain. The increasing convergence of the applications of autonomous vehicles, smart cities, cognitive wireless networks, and industrial automation have for the first time stressed the necessity of unifying these disjoint functionalities as the demands of the applications are now sophisticated in terms of the requirements for high speed communication along with the real-time awareness of the situation. \cite{zhang2021overview} The convergence is now becoming a reality due to not only the needs of the applications, but also due to the maturity of the modern techniques of signal processing and their integration with hardware.

\subsection{Definition and Evolution}
JCAS is the first unified conception of design and deployment of systems that make use of integrated spectral, hardware, and/or signal processing resources that are common for data communication and environment sensing functions \cite{sheemar2023full}. The core idea is to employ the dual purpose of RF signals: communication systems, in ignoring sensing, regard the environment as a hindrance (e.g., multipath, shadowing), while sensing systems use the same propagation to understand the environment. The evolution of JCAS can be traced through three major phases:
\begin{itemize}

    \item \textbf{First Signs of Alignement in Frequency Usage}: The first sign of alignment (or convergence) was in the early instances of spectrum collision environments where radar and comm systems were operating in overlapping frequency bands. The first signs of synergy were the more advanced techniques of mitigating interference \cite{chiriyath2017radar}.

    \item \textbf{Awareness and Reuse of Waveforms}: The possibility of a two-way street was first recognized in the use of comm waveforms and basic radar functions, in the case of OFDM and spread spectrum \cite{bicua2019multicarrier}. Likewise, radar signals were shown to have the ability to potentially carry a data payload, and thus described using an information theoretic approach \cite{temiz2023radar}.

    \item \textbf{Integrated Design Systems}: The designs of integrated systems in modern JCAS R\&D, particularly co-design, where the waveform, protocol, and hardware integration are balanced to optimize convergence in the spectrum to enable the simultaneous supporting functions \cite{sheemar2023full}. This co-design strategy is associated with large gains in spectral efficiency, reductions in latency and energy consumption, and increases in robustness of the system \cite{sheemar2025holographic}. Rudimentary JCAS capabilities have already been demonstrated in systems like the IEEE 802.11ad/ay (WiGig) and 5G NR due to their capabilities in wide bandwidth and advanced beamforming.

\end{itemize}

\begin{table*}[t]
\centering 
\caption{Comparison of JCAS Waveform Usages}
\label{table:JCAS_comparison}
\renewcommand{\arraystretch}{1.25}
\setlength{\tabcolsep}{6pt}

\begin{tabular}{|p{3.3cm}|p{4.7cm}|p{4cm}|p{4cm}|}
\hline
\rowcolor{HeaderRed}
\textbf{Aspect} & \textbf{Communication-centric JCAS} & \textbf{Sensing-centric JCAS} & \textbf{Dual-function JCAS} \\
\hline

\rowcolor{BodyBlue}
\textbf{Waveform Origin} & Based on communication signals (e.g., OFDM, mmWave) & Based on radar/sensing pulses (e.g., FMCW, chirp) & Custom-designed for both sensing and communication \\
\hline

\rowcolor{BodyBlue}
\textbf{Primary Goal} & Maximize communication rate; reuse waveform for sensing & Maximize sensing resolution; embed data into radar waveform & Jointly optimize communication and sensing performance \\
\hline

\rowcolor{BodyBlue}
\textbf{Use Cases} & Mobile networks, 5G NR, legacy systems & Automotive radar, UAV navigation, tracking & JCAS base stations, 6G, smart cities, autonomous driving \\
\hline

\rowcolor{BodyBlue}
\textbf{Communication Strategy} & Use existing modulation (e.g., QAM) & Modulate radar pulses (e.g., ASK, PPM, phase coding) & Joint waveform/precoder design for rate--sensing trade-off \\
\hline

\rowcolor{BodyBlue}
\textbf{Sensing Strategy} & Delay/Doppler estimation via reflected comm signal & High-resolution detection using radar signal processing & Joint estimation with optimized waveform and beamforming \\
\hline

\rowcolor{BodyBlue}
\textbf{Optimization Objective} & Maximize throughput with minimal sensing loss & Maximize sensing accuracy with minimal data loss & Multi-objective optimization for dual purpose \\
\hline

\rowcolor{BodyBlue}
\textbf{Complexity} & Low (backward compatible) & Medium (requires sensing-side modifications) & High (requires joint design and more coordination) \\
\hline

\rowcolor{BodyBlue}
\textbf{Hardware Requirement} & Communication transceiver compatible & Radar-capable front-end required & Adaptable front-end and baseband processing required \\
\hline
\end{tabular}
\end{table*}

\subsection{Different levels of Optimization}
The main idea for JCAS rest in what we call joint design, or  co-optimization of communication and sensing functions at different levels of the system stack.  While traditional methodologies segregate sensing and communication as distinct subsystems, in joint design we seek to unify waveform, hardware, signal processing, and network protocols for the dual role to be achieved in an optimized manner and at the same time. These can be categorized at different levels as:

\begin{itemize}
    \item At the \emph{waveform level}, signal structure plays an important role. For instance, orthogonal frequency division multiplexing (OFDM) is a candidate for communication systems due to its ability to cope with multipath issues, along with flexible frequency distribution \cite{wu2023joint}. OFDM is also advantageous for radar processing, as its frequency diversity and cyclic prefix facilitate precise range and Doppler estimation. Automotive radar also uses a Waveform like frequency modulated continuous wave (FMCW) \cite{venon2022millimeter} which is being modified to transmit data as well \cite{wang2025coordinated}. Here, the problem is to optimize the trade-offs pertaining to a high peak-to-average power ratio (PAPR) and low sidelobe levels for sensing, contrasted with high spectral efficiency and low error rate for communication.

    \item From a \emph{signal processing perspective}, joint design has to integrate data demodulation and target parameter estimation \cite{zhang2021overview}. Matched filtering, beamforming, and compressive sensing need to be reconstructed to function under joint performance metrics. Consider, for instance, a vehicle. In this context, beam alignment techniques should be designed to not only optimize the quality of the communication link, but also to provide adequate angular resolution for the localization of the target. In the same manner, the sensing of clutter and multipath can be used to enhance the reliability of communication by introducing spatial diversity.

    \item On the \emph{hardware level}, the joint design will need to accommodate shared antennas, RF front-ends, and possibly even baseband processing chains. This is beneficial for size, weight, power, and cost, all of which are critical for platforms such as drones, vehicles, and smartphones. However, it creates a number of challenges to the system design in order to ensure that there is no mutual interference between the sensing and communication functions, particularly in full-duplex or simultaneous operations \cite{sheemar2023full}.

    \item The complexity of joint design escalates on the \emph{protocol and network layer}. At this level, the combined challenges of scheduling resources (power, time, frequency, and space) and maintaining the desired communication quality of service (QoS) alongside sensing coverage and resolution pose additional challenges. New optimization problems and trade-offs are introduced. For instance, in a multi-user context, the base station has to make decisions on which beams to afford in order to service users and to surveil areas of interest in the ecosystem.

    \item Additionally, \emph{context-awareness and learning-based} methods are becoming important facilitators of integrated design \cite{farzanullah2025beam}. Utilizing historical information, the context of the environment, and predictive models (for instance, machine learning), JCAS systems are able to flexibly manage and shift focus to either communication or sensing depending on the current operational situation. In a highway example, where the road geometry is known, the sensing task can be simplified so that more resources can be devoted to mission-critical communication.
\end{itemize}

\subsection{Classification of JCAS Systems}
In the following subsection, we provide an overview of how JCAS can be classified at different levels in terms of hardware, signals, and system architecture, as discussed in the following. 

\subsubsection{Based on Integration Level} JCAS can be categorized into three architectures based on integration level, i.e. coexistance-based JCAS, cooperative JCAS and joint JCAS.
\begin{itemize}
    \item \emph{Coexistence-Based systems:} In this architecture, the sensing and communication systems operate independently while sharing the same spectrum. There is minimal coordination between the two and interference mitigation strategies are required to ensure reliable performance.
    \item \emph{Cooperative systems:} Separate sensing and communication systems collaborate via coordination protocols or data exchange. Although hardware may remain separate, joint optimization improves performance across both functionalities.
    \item \emph{Joint/Unified systems:} A single system jointly performs both sensing and communication using the same hardware and waveform resources. This is the most integrated form, achieving high spectral and energy efficiency.
\end{itemize}

\begin{figure*}[!t] 
\centering
\begin{tikzpicture}[
    node distance = 1.5cm,
    bs/.style = {draw, rectangle, minimum width=1.5cm, minimum height=0.8cm, fill=blue!20, align=center},
    target/.style = {circle, draw, fill=red!20, minimum size=0.8cm},
    user/.style = {circle, draw, fill=green!20, minimum size=0.8cm},
    arrow/.style = {->, >=stealth, thick},
    commarrow/.style = {->, >=stealth, thick, dashed, blue},
    label/.style = {font=\footnotesize, align=center}
]

\node[bs] (mono) at (0,0) {Tx + Rx};
\node[target] (target1) at (4,0) {Target};
\node[user] (user1) at (2,1.5) {User};
\draw[arrow] (mono) -- node[above, label, pos=0.7] {Sensing} (target1);
\draw[commarrow] (mono) -- node[left, label] {Comm} (user1);
\draw[arrow, dashed] (target1) -- node[below, label] {Echo} (mono);
\node[below=0.5cm of mono, label] {\textbf{(a) Monostatic JCAS}};

\node[bs] (tx) at (5,-3) {Tx};
\node[bs] (rx) at (8,-3) {Rx};
\node[target] (target2) at (6.5,-1.5) {Target};
\node[user] (user2) at (7,-4.5) {User};
\draw[arrow] (tx) -- node[above left, label] {Sensing} (target2);
\draw[commarrow] (tx) -- node[right, label] {Comm} (user2);
\draw[arrow, dashed] (target2) -- node[above right, label] {Echo} (rx);
\node[below=1cm of tx, label] {\textbf{(b) Bistatic JCAS}};

\node[bs] (tx1) at (9,0) {Tx1};
\node[bs] (tx2) at (11,-1.5) {Tx2}; 
\node[bs] (rx1) at (13.5,1.5) {Rx1}; 
\node[bs] (rx2) at (15,0) {Rx2}; 
\node[target] (target3) at (12,0) {Target}; 
\node[user] (user3a) at (9,2) {User}; 
\node[user] (user3b) at (14,-3) {User}; 
\draw[arrow] (tx1) -- (target3);
\draw[arrow] (tx2) -- (target3);
\draw[commarrow] (tx1) to[out=90,in=-90] node[left, pos=0.3] {Comm} (user3a);
\draw[commarrow] (tx2) -- node[right, pos=0.4] {Comm} (user3b); 
\draw[arrow, dashed] (target3) -- node[above left, pos=0.7] {} (rx1); 
\draw[arrow, dashed] (target3) -- (rx2);
\node[below=2.5cm of target3, label] {\textbf{(c) Multistatic JCAS}};

\end{tikzpicture}
\caption{JCAS architectures showing both sensing and communication: (a) Monostatic, (b) Bistatic, and (c) Multistatic configurations. Solid arrows show sensing signals, dashed arrows show echoes, and blue dashed arrows show communication.}
\label{fig:JCAS_architectures}
\end{figure*}
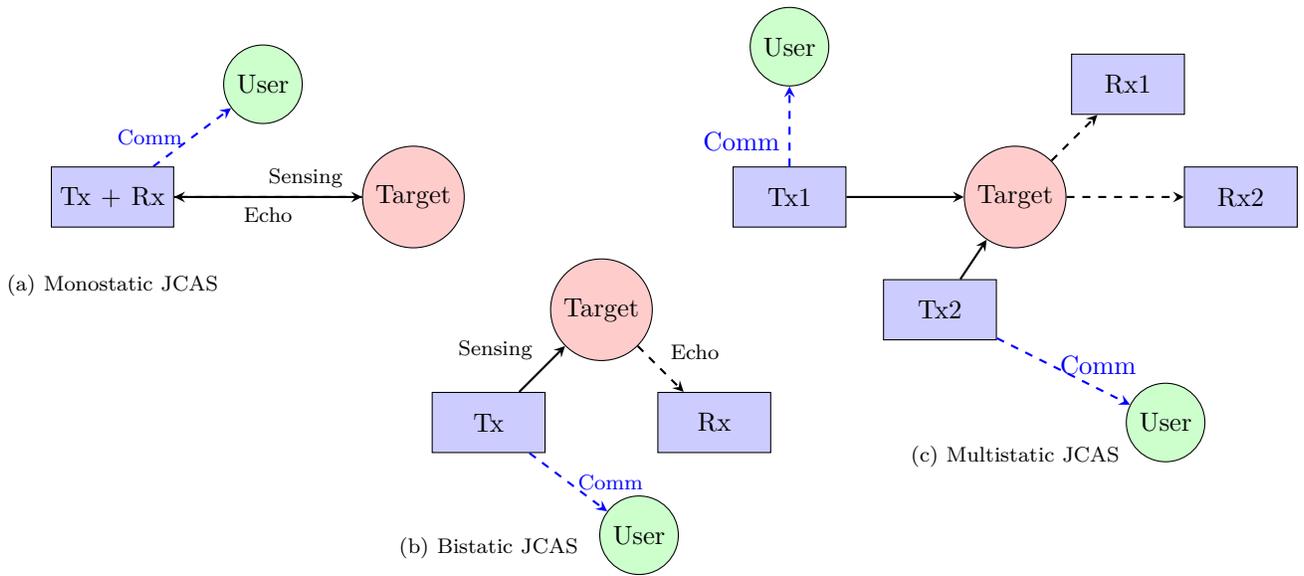
\subsubsection{Based on Waveform Usage}
JCAS systems can be further classified based on the nature and purpose of waveform utilization into three categories, as illustrated in Table \ref{table:JCAS_comparison}, and discussed in detail in the following.

\noindent\textit{a) Communication-Centric JCAS:}  
Communication-centric JCAS takes advantage of existing communication waveforms such as OFDM, beamform signals, and massive MIMO transmissions to enable sensing functions. This approach is attractive due to its compatibility with current communication standards, allowing sensing to be incorporated primarily through enhanced receiver-side processing with minimal changes to the transmitter \cite{naoumi2025high}. Sensing is achieved by processing reflected and Doppler-shifted replicas of the transmitted communication signal to estimate environmental parameters such as range, velocity, and angle \cite{cheng2022integrated}. The wide bandwidth, multicarrier structure, and availability of pilot signals in modern communication systems facilitate such estimation. However, since the waveform is not explicitly designed for sensing, achievable sensing resolution and ambiguity suppression are generally limited. As a result, communication-centric JCAS is best suited for applications where communication performance is prioritized and sensing is performed opportunistically.

\noindent\textit{b) Sensing-Centric JCAS:}  
Sensing-centric JCAS is based on waveforms originally designed for radar or active sensing, such as FMCW \cite{venon2022millimeter}, chirp \cite{fink2015comparison}, or pulse-Doppler signals, which are adapted to carry communication data \cite{ahmed2023sensing}. Information is embedded into the sensing waveform through modulation techniques that preserve sensing performance, including phase, amplitude, or pulse position modulation. This approach enables high-resolution sensing with accurate range and velocity estimation, making it particularly suitable for short-range and safety-critical applications such as automotive radar and UAV navigation. However, because the waveform is primarily optimized for sensing, the achievable communication data rate is often limited. Consequently, sensing-centric JCAS is most appropriate in scenarios where sensing accuracy is the dominant requirement and communication plays a secondary role.

\noindent\textit{c) Dual-Function JCAS:}  
Dual-function JCAS employs waveforms that are explicitly designed to simultaneously support both sensing and communication objectives \cite{zhang2021enabling}. Rather than favoring one function over the other, the waveform and associated transmission parameters are jointly optimized to balance sensing accuracy and communication performance \cite{sheemar2025joint}. This joint design is commonly formulated as a multi-objective optimization problem, allowing the system to flexibly trade communication data rate against sensing estimation error depending on application requirements. Dual-function JCAS offers the highest degree of adaptability and performance synergy, but typically entails increased computational complexity and more stringent hardware requirements. As such, it is widely regarded as a foundational approach for native JCAS operation in future 6G networks.

\subsubsection{Based on System Architecture}
Similar to other categories, architecture-wise, JCAS can be classified into three categories depending on the position of the receiver, as in the following \cite{sheemar2023full}, also highlighted in Figure \ref{fig:JCAS_architectures}.  

\begin{itemize}
    \item \emph{Monostatic JCAS:} Both transmitter and receiver are positioned at the same location \cite{lu2022degrees,keskin2025fundamental}. This setup is commonly used in base stations, autonomous vehicles, and access points.
    \item{Bistatic JCAS:} The transmitter and receiver are placed at different physical locations \cite{zhao2024joint,naoumi2024complex}. This configuration allows for broader coverage and is useful for remote sensing applications.
    \item{Multistatic/Networked JCAS:} Multiple transmitters and receivers cooperate in a distributed manner to perform JCAS over a wider area, improving spatial diversity and robustness \cite{hanle1986survey}.
\end{itemize}
 
\subsection{Applications of JCAS}
JCAS systems can be tailored to meet the specific needs of different application scenarios by jointly supporting perception and data exchange functionalities. The following are key domains where JCAS has transformative potential in 6G networks.
\subsubsection{Vehicular Networks (V2X)} 
    In vehicular-to-everything (V2X) communications, JCAS can be employed to simultaneously support high-speed data transmission and real-time environment perception \cite{decarli2024performance}. For example, a vehicle equipped with JCAS-enabled transceivers can exchange safety-critical messages with nearby vehicles (V2V), road infrastructure (V2I), and pedestrians (V2P), while also detecting objects (e.g., pedestrians, other vehicles) and estimating their speed, range, and direction using the same waveform \cite{giovannetti2023target}. This enables advanced driver assistance systems (ADAS), autonomous driving, and collision avoidance in dynamic road environments.
\subsubsection{Industrial Internet of Things (IIoT)}
    In smart manufacturing environments, JCAS enables real-time control and monitoring of robotic arms, automated guided vehicles (AGVs), and machinery \cite{wu2021otfs}. JCAS allows these devices to perceive their surroundings (e.g., obstacles, human workers) while simultaneously receiving control commands or transmitting sensor data. JCAS ensures ultra-low latency and high reliability, which are essential for time-critical industrial automation and safety assurance.
    
\subsubsection{Indoor Mapping and Localization}
    In complex indoor environments such as airports, shopping malls, and hospitals, JCAS facilitates accurate user localization, motion tracking, and map generation alongside wireless communication \cite{nemati2022toward}. For instance, access points equipped with JCAS can sense user locations or gestures while providing Wi-Fi or 6G connectivity. This dual functionality improves navigation, enhances user experiences, and supports context-aware services such as AR/VR or asset tracking.

\subsubsection{Non-Terrestrial Networks (NTNs)}
    In satellite and aerial communication systems, JCAS enables communication and Earth observation capabilities jointly \cite{sheemar2025joint}. A satellite or UAV can transmit high-data-rate signals to ground terminals while simultaneously sensing terrestrial features (e.g., land use, environmental conditions, or disaster zones). This reduces payload redundancy and is highly beneficial for remote sensing, environmental monitoring, and integrated communication services in hard-to-reach areas.

\begin{figure*}[!t]
\centering
\includegraphics[width=0.8\textwidth,height=6cm]{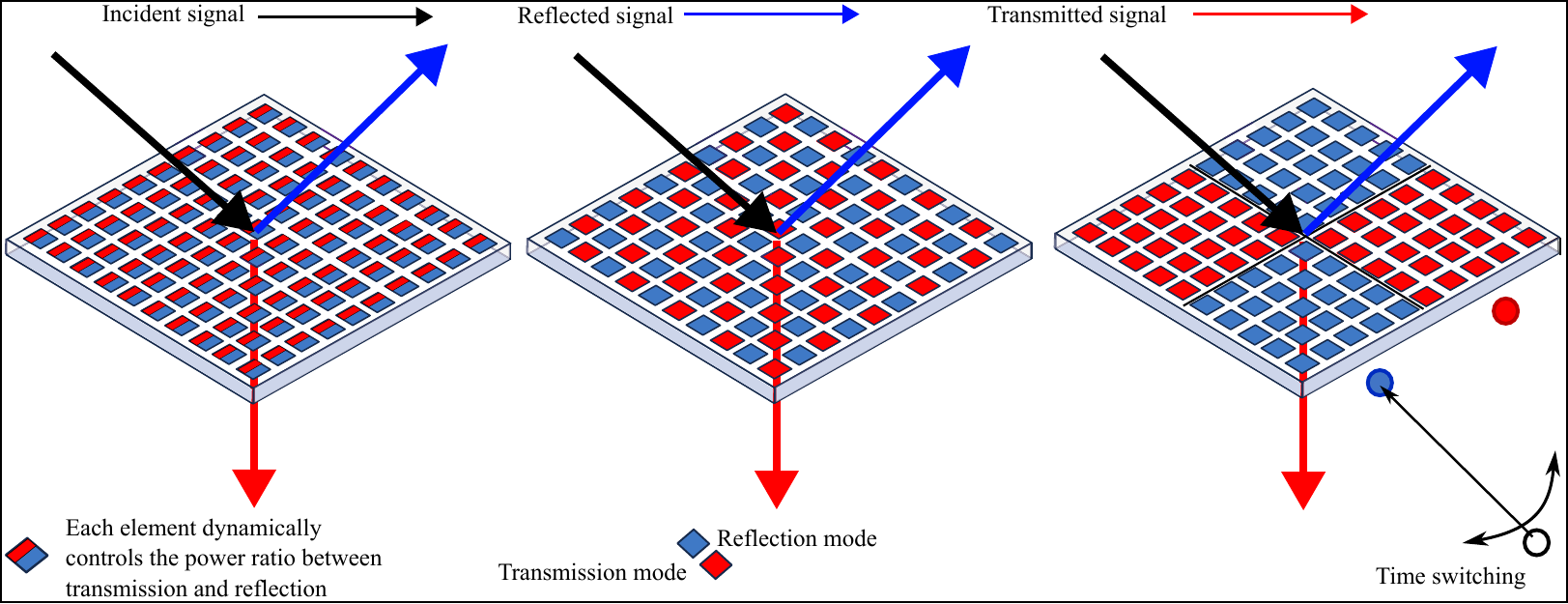}
\caption{Operating protocols of STAR-RIS: (a) ES STAR-RIS, (b) MS STAR-RIS and (c) TS STAR-RIS.}
\label{STAR1}
\end{figure*}
\subsubsection{Smart Healthcare and Surveillance}
    In healthcare applications, JCAS facilitates non-invasive monitoring of vital signs (e.g., respiration rate, heart rate) while also providing data connectivity \cite{al2024comprehensive}. For example, radar-based JCAS systems can detect micro-movements associated with breathing or falls, while transmitting the patient's health data to a cloud server or a nearby care center. In public surveillance, JCAS can detect suspicious movements or activities while streaming video and sensor data to control centers, improving safety and  efficiency.

\section{What is STAR-RIS?}
In this section, we cover the basics of the STAR-RIS technology.
We first present the foundational signal model for STAR-RIS and introduce three operating protocols to integrate these surfaces into wireless communication systems. We also discuss the advantages and disadvantages of each protocol. Subsequently, we discuss the passive and active STAR-RIS architectures, as well as their applications in 6G networks. 

\subsection{STAR-RIS Signal Model}

STAR-RISs are designed to independently control both transmitted and reflected signals, offering additional degrees of freedom (DoFs) that can be leveraged. For a STAR-RIS with $N$ elements, let $x_{n}$ represent the signal incident on the $n-$th element. After being adjusted by the corresponding transmission and reflection coefficients, the signals transmitted and reflected by the $n$th element are given by $\left(\sqrt{\alpha_{n}^{t}}e^{j\phi_{n}^{t}}\right)x_{n}$ and $\left(\sqrt{\alpha_{n}^{r}}e^{j\phi_{n}^{r}}\right)x_{n}$, respectively. Here, $\sqrt{\alpha_{n}^{t}}$, $\sqrt{\alpha_{n}^{r}} \in [0,1]$ and $\phi_{n}^{t}, \phi_{n}^{r} \in [0,2\pi)$ denote the amplitude and phase shift adjustments for transmission and reflection, respectively. These adjustments can generally be chosen independently. However, the amplitude coefficients for transmission and reflection must adhere to the energy conservation principle, which means $\alpha_{n}^{t} + \alpha_{n}^{r} = 1$. This allows each STAR-RIS element to operate in three modes: full transmission (T mode), full reflection (R mode), or simultaneous transmission and reflection (T\&R mode).

\subsection{STAR-RIS Operating Protocols}
Considering the signal model discussed in the previous section, we discuss three practical protocols for operating STAR-RISs in wireless networks: 1) energy splitting (ES), 2) Mode Switching (MS), and 3) time switching (TS), as illustrated in Fig. \ref{STAR1}.

\subsubsection{ES STAR-RIS}

In the ES protocol, all elements of the STAR-RIS operate in the T\&R mode, as shown in Fig. \ref{STAR1}(a). The incident signal on each element is divided into transmitted and reflected components with different energy levels. The amplitude and phase shift coefficients for transmission and reflection can be jointly optimized to achieve various design objectives in wireless networks.

\subsubsection{MS STAR-RIS}

In the MS protocol, the elements of STAR-RIS are divided into two groups: one group operates in the T mode, and the other operates in the R mode, as shown in Fig. \ref{STAR1}(b). This protocol can be viewed as combining a conventional reflecting-only RIS and a transmitting-only RIS, each with reduced sizes. The mode selection for each element and the corresponding phase shift coefficients can be jointly optimized. Although this "on-off" type protocol is straightforward to implement, it generally cannot match the transmission and reflection gains of ES, as only a subset of elements is used for each function.

\subsubsection{TS STAR-RIS}
In the TS protocol, all elements of the STAR-RIS periodically switch between the T mode and the R mode in orthogonal time slots, known as the T period and the R period, as illustrated in Fig. \ref{STAR1}(c). This is akin to adjusting "venetian blinds" over time. The time allocated to transmission and reflection can be optimized to balance the communication quality on both sides of the RIS. Unlike ES and MS, TS allows for independent optimization of transmission and reflection coefficients for a given time allocation. However, the need for periodic switching imposes strict time synchronization requirements, increasing implementation complexity compared to ES and MS. The optimization variables of the three protocols mentioned above, together with the advantages and disadvantages, are concisely stated in Table \ref{tab:protocols}.

\begin{table}[!h]
\centering
\caption{Summary of ES, MS, and TS Protocols}
\label{tab:protocols}
\renewcommand{\arraystretch}{1.2}
\setlength{\tabcolsep}{7pt}
\begin{tabular}{|p{1cm}|p{2.9cm}|p{3.5cm}|}
\hline
\rowcolor{HeaderRed}
\textbf{Protocol} & \textbf{Optimization Variables} & \textbf{Advantages and Disadvantages} \\ \hline

\rowcolor{BodyBlue}
ES & Amplitude and phase for transmission/reflection & High flexibility and performance; higher optimization complexity \\ \hline

\rowcolor{BodyBlue}
MS & Mode selection (T or R) and phase shifts & Simple implementation; reduced degrees of freedom and gain \\ \hline

\rowcolor{BodyBlue}
TS & Time allocation for T and R (plus phases) & Decoupled design across time slots; requires time synchronization \\ \hline
\end{tabular}
\end{table}

\begin{figure*}[!t]
\centering
\includegraphics[width=0.8\textwidth]{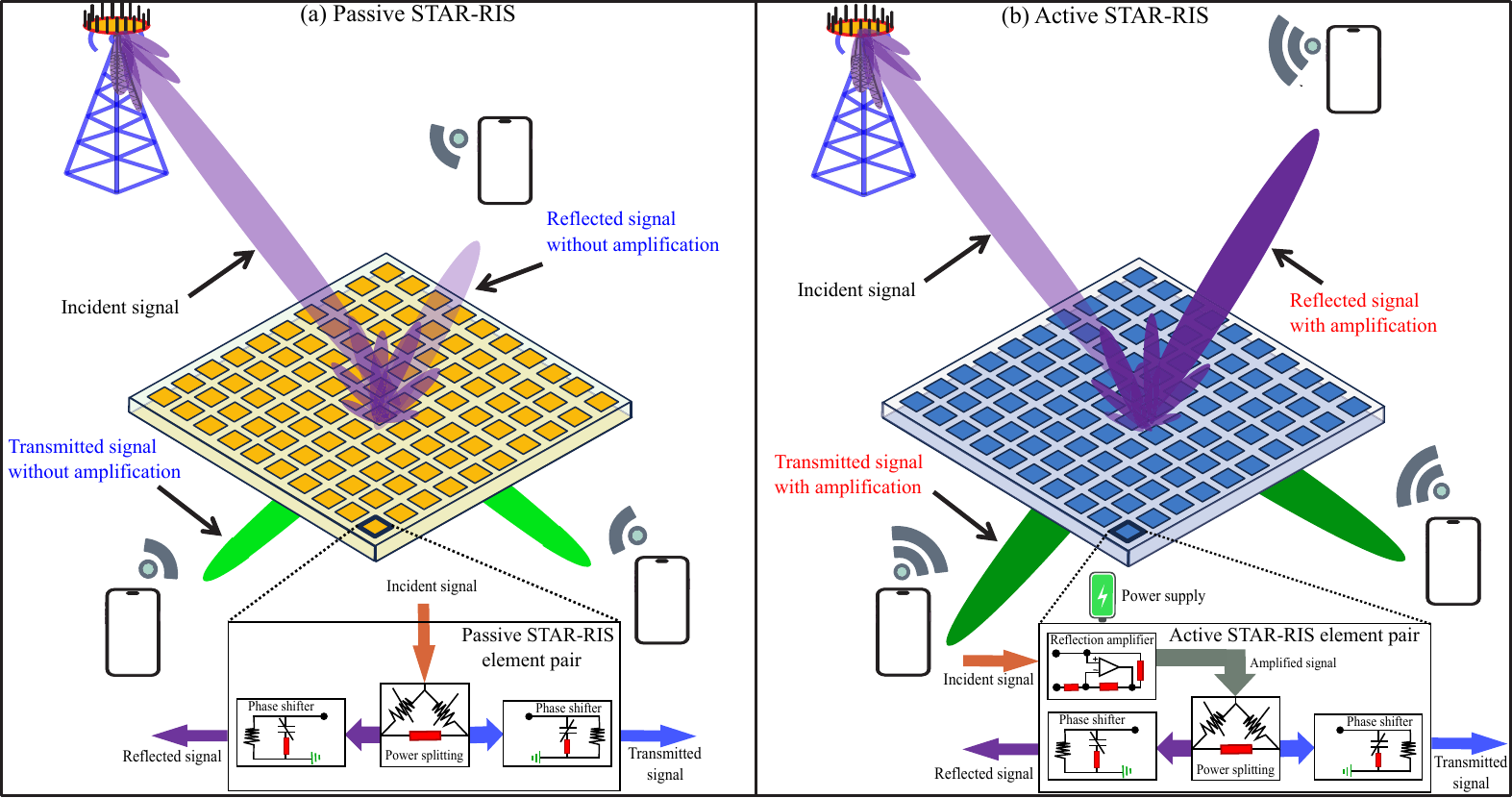}
\caption{Architectures of STAR-RIS: (a) Passive STAR-RIS, (b) Active STAR-RIS.}
\label{STAR2}
\end{figure*}

\subsection{Passive STAR-RIS}
A passive STAR-RIS operates without the need for active components, such as amplifiers or signal generators, relying solely on the reconfigurable properties of its elements to manipulate incident signals \cite{zhu2025star}. Unlike active RISs, which require external power sources to amplify or regenerate signals, passive STAR-RISs achieve signal manipulation through the careful design of their metasurface elements, which can independently control the phase and amplitude of transmitted and reflected waves. This passive operation makes STAR-RISs highly energy-efficient and cost-effective, as they do not consume additional power for signal processing \cite{10792983}. However, the lack of active components limits their ability to compensate for signal attenuation over long distances, making them more suitable for short- to medium-range applications. Despite this limitation, passive STAR-RISs are highly attractive for deployment in scenarios where energy efficiency and low maintenance are critical, such as in smart buildings, indoor communication systems, and urban environments with dense user populations. Their ability to simultaneously transmit and reflect signals in a full-space coverage area further enhances their utility in next-generation wireless networks.

\subsection{Active STAR-RIS}
An active STAR-RIS incorporates active components, such as amplifiers, signal generators, or integrated circuits, to improve its signal manipulation capabilities \cite{zhu2025star}. Unlike passive STAR-RISs, which rely solely on the reconfigurable properties of their metasurface elements, active STAR-RISs can amplify or regenerate incident signals, thereby compensating for signal attenuation over long distances. This makes them particularly suitable for applications that require extended coverage or high signal strength, such as in rural areas, large indoor spaces, or environments with significant signal blockage \cite{10753360}. Inclusion of active components allows for more sophisticated control over transmitted and reflected signals, allowing advanced functionalities like beamforming, interference cancellation, and dynamic adaptation to changing channel conditions. However, the use of active components increases the power consumption and cost of the STAR-RIS, making it less energy-efficient compared to its passive counterpart. Despite these drawbacks, active STAR-RISs are highly advantageous in scenarios where performance and flexibility are prioritized over energy efficiency, such as in high-capacity communication systems, millimeter-wave (mmWave) or terahertz (THz) networks \cite{kurma2024resource}, and mission-critical applications such as emergency communications or industrial automation.

\subsection{STAR-RIS Features}
In the following, we highlight the key-features of STAR-RIS that offer several advantages over conventional RIS technologies \cite{P-RISOJCOM}.

\subsubsection{Full-Space Coverage} 
Unlike conventional RIS, which only reflects signals, STAR-RIS ensures complete 360-degree coverage by simultaneously transmitting and reflecting signals. This enhances communication reliability in multi-user and multi-path environments.
\subsubsection{Enhanced Adaptive Beamforming} STAR-RIS can dynamically adjust the transmission and reflection coefficients to optimize the signal-to-noise ratio (SNR) for assisted users. STAR-RIS enables the joint optimization of reflection and transmission parameters, making it adaptable to various communication scenarios, including space-air-ground integrated networks.
\subsubsection{Improved spectral and energy efficiency} STAR-RIS can improve signal utilization through intelligent reconfiguration in the full space, resulting in improved spectral efficiency. By intelligently controlling the propagation environment without requiring additional power-hungry RF chains, STAR-RIS contributes to green communication networks while providing enhanced degrees of freedom for maximizing spectral and energy efficiency.
\subsubsection{Stronger Connectivity in NLOS Scenarios} STAR-RIS is particularly useful in non-line-of-sight (NLOS) environments, such as urban settings or indoor deployments, where traditional RIS may not provide adequate coverage, or only partially cover the area of interest. In addition to the aforementioned key features, additional features are listed in Table \ref{tab:RIS_comparison}, compared to conventional RIS technologies.

\begin{table*}[!t]
\centering
\caption{Comparison of Reflective RIS, Transmissive RIS, and STAR-RIS}
\label{tab:RIS_comparison}
\renewcommand{\arraystretch}{1.2}
\setlength{\tabcolsep}{7pt}


\begin{tabular}{|p{2.8cm}|p{3.9cm}|p{4.0cm}|p{5.3cm}|}
\hline
\rowcolor{HeaderRed}
\textbf{Feature} & \textbf{Reflective RIS} & \textbf{Transmissive RIS} & \textbf{STAR-RIS} \\
\hline

\rowcolor{BodyBlue}
\textbf{Signal Manipulation} & Reflection only & Transmission only & Simultaneous reflection and transmission \\
\hline

\rowcolor{BodyBlue}
\textbf{Coverage Region} & One side of the RIS ($180^\circ$) & Opposite side of the RIS ($180^\circ$) & Both sides of the RIS ($360^\circ$) \\
\hline

\rowcolor{BodyBlue}
\textbf{Energy Utilization} & Entire signal is reflected & Entire signal is transmitted & Adjustable split between reflection and transmission \\
\hline

\rowcolor{BodyBlue}
\textbf{Deployment Flexibility} & Limited to reflective deployment & Limited to transmissive deployment & More flexible across diverse deployments \\
\hline

\rowcolor{BodyBlue}
\textbf{Operational Modes} & Reflection mode & Transmission mode & ES, MS, and TS protocols \\
\hline

\rowcolor{BodyBlue}
\textbf{Power Efficiency} & High (passive) & High (passive) & Mode-dependent (typically moderate to high) \\
\hline

\rowcolor{BodyBlue}
\textbf{Energy Efficiency} & High & High & Mode- and allocation-dependent \\
\hline

\rowcolor{BodyBlue}
\textbf{Cost} & Low (simple structure) & Low (simple structure) & Higher due to dual-mode circuitry/control \\
\hline

\rowcolor{BodyBlue}
\textbf{Complexity} & Low (passive control) & Low (similar to reflective RIS) & Higher due to dual-mode operation and control \\
\hline

\rowcolor{BodyBlue}
\textbf{Application Scenarios} & Indoor coverage extension, wireless backhauling & Transparent surfaces, smart-glass-like deployments & 6G, ISAC, smart cities, and space--air--ground networks \\
\hline
\end{tabular}
\end{table*}

 \subsection{STAR-RIS Applications}
Unlike conventional RIS that operate only in reflection or transmission mode, STAR-RIS simultaneously supports both functions, enabling full-space signal manipulation. This fundamental capability allows STAR-RIS to unlock new functionalities and significantly enhance performance across a wide range of wireless applications, as outlined below.

\subsubsection{Non-Orthogonal Multiple Access (NOMA)}
In contrast to conventional RIS, which can only assist users located on one side of the surface, STAR-RIS enables simultaneous service of users in both the transmission and reflection regions \cite{11202653}. This full-space coverage significantly increases the user pairing pool in NOMA systems, allowing more flexible user grouping and improved power-domain multiplexing. As a result, STAR-RIS achieves higher spectral efficiency and supports massive connectivity more effectively than reflecting-only or transmitting-only RIS, which inherently limit user access to half-space coverage.

\subsubsection{Coordinated Multi-Point (CoMP) Communication}
Conventional RIS can only enhance links within or outside a cell depending on their deployment orientation. STAR-RIS overcomes this limitation by enabling dual-sided beamforming, allowing a single surface to simultaneously assist multiple base stations and users located in different cells \cite{umer2023performance}. This capability improves interference coordination, especially for cell-edge users, and enhances network-wide throughput without requiring additional transmit power or active relaying.

\subsubsection{Physical Layer Security (PLS)}
Traditional RIS-based security solutions are constrained by the relative positions of legitimate users and eavesdroppers with respect to the surface. STAR-RIS removes this restriction by independently controlling transmitted and reflected signals, allowing simultaneous enhancement of legitimate links and suppression of eavesdropping channels on both sides of the surface \cite{liu2023exploiting}. This location-agnostic security control provides stronger and more flexible physical-layer security compared to conventional RIS architectures.

\subsubsection{Indoor Localization and Sensing}
Conventional RIS improves indoor localization mainly by enhancing reflected paths. STAR-RIS further improves positioning accuracy by generating controllable signal paths in both directions, effectively enriching the multipath environment \cite{he2023star}. This full-space wave control increases spatial diversity and resolution, enabling more accurate indoor localization, motion detection, and environment sensing in smart factories, warehouses, and smart buildings.

\subsubsection{Simultaneous Wireless Information and Power Transfer (SWIPT)}
In SWIPT systems, conventional RIS must prioritize information decoding or energy harvesting users within the same spatial region. STAR-RIS enables simultaneous support of information and energy receivers located on opposite sides of the surface by appropriately splitting and shaping the transmitted and reflected signals \cite{zhu2023robust}. This dual-sided operation improves energy efficiency and flexibility, making STAR-RIS particularly attractive for battery-constrained IoT networks.

\subsubsection{mmWave and THz Communications}
At mmWave and THz frequencies, conventional RIS can only mitigate blockages for users positioned in the reflection region. STAR-RIS extends blockage mitigation to both sides of the surface by creating virtual line-of-sight links via transmission and reflection \cite{yan2024wideband}. This capability significantly improves coverage continuity, reliability, and link robustness in dense urban and indoor environments where high-frequency signals are highly directional and vulnerable to obstruction.

\subsubsection{UAV and Vehicular Networks}
In high-mobility scenarios, conventional RIS provides limited assistance due to its half-space coverage and fixed orientation constraints. STAR-RIS enables the dynamic beamsteering toward moving UAVs \cite{zhang2022joint} and vehicles on both sides of the surface \cite{aung2023deep}, facilitating connectivity and joint localization–communication support. This reduces signaling overhead and improves reliability in vehicle-to-infrastructure (V2I) and drone-assisted communication networks.

\subsubsection{Visible Light Communication (VLC)}
For VLC and hybrid RF–optical systems, conventional reflective surfaces primarily enhance illumination or communication in a single direction. STAR-RIS extends this capability by enabling controllable transmission and reflection of visible light, improving coverage uniformity, and enabling joint lighting and communication optimization \cite{maraqa2023optical}. Although less explored than RF applications, STAR-RIS offers promising opportunities for smart indoor environments and dual-use lighting–communication systems.

\subsubsection{JCAS}
For JCAS systems, conventional RIS can only assist in detecting or communicating targets located in the same half-space. STAR-RIS enables concurrent optimization of communication and sensing functions in different spatial regions by splitting energy between transmission and reflection \cite{LIU-JIOT}. This capability allows STAR-RIS to jointly improve data rates and sensing accuracy (e.g., target detection and tracking), making it particularly suitable for vehicular networks, smart transportation, and UAV-based sensing applications.

\section{STAR-RIS Enabled JCAS: Recent Advances}

Based on the discussion in the preceding sections, it is evident that the convergence of communication and sensing in 6G networks requires advanced technologies that are low-cost, energy-efficient, open and modular to support JCAS. Among these, STAR-RIS has emerged as a compelling candidate for RIS-assisted JCAS due to its unique capability to achieve full-space electromagnetic coverage through simultaneous transmission and reflection. This section critically analyzes recent developments in STAR-RIS-enabled JCAS frameworks, focusing on architectural innovations, optimization strategies, waveform design, and application-specific implementations.

\begin{table*}[!t]
\centering
\caption{Summary of architectural designs and optimization strategies in STAR-RIS enabled JCAS.}
\label{tab:IV_A_summary}
\renewcommand{\arraystretch}{1.08}
\setlength{\tabcolsep}{5.2pt}


\begin{tabular}{|p{0.7cm}|p{2.2cm}|p{2cm}|p{2.7cm}|p{3.0cm}|p{1.5cm}|p{3.2cm}|}
\hline
\rowcolor{HeaderRed}
\textbf{Ref.} & \textbf{Scenario/Setup} & \textbf{Architecture/ Protocol} & \textbf{Objective} & \textbf{Optimized Variables} & \textbf{Method} & \textbf{Metrics/ Notes} \\
\hline

\rowcolor{BodyBlue}
\cite{ZehraHybrid} 2025 &
Multi-user/ multi-target JCAS &
Hybrid STAR-RIS  &
Joint comm--sensing design (SINR-based) &
T/R coefficients, BS beamforming &
QCQP $\rightarrow$ SDR (CVX) &
CRB (AoD), comm SINR; highlights full-space gain \\
\hline

\rowcolor{BodyBlue}
\cite{P-RISOJCOM} 2024 &
Multi-user DFRC with metasurface assistance &
Hybrid STAR-RIS, P-RIS &
Max WSR under sensing constraints &
BS beamforming, RIS phase/coefficients &
WMMSE, AO &
WSR/SE, sensing constraints; assumes ideal CSI \\
\hline

\rowcolor{BodyBlue}
\cite{11059504} 2025 &
Multi-user/ single-target JCAS &
ES protocol STAR-RIS &
Max radar SINR with comm SINR constraints &
BS beamforming, STAR-RIS coeffs &
FP, BCD, SDR &
Radar SINR, comm SINR; strong non-convex coupling \\
\hline

\rowcolor{BodyBlue}
\cite{10973322} 2025 &
Single-user/ single-target JCAS &
Statistical CSI model &
Tradeoff: ergodic rate vs sensing SCNR &
BS beamforming, STAR-RIS phases &
AO/ stochastic design &
Ergodic rate, SCNR; improves realism vs perfect CSI \\
\hline

\rowcolor{BodyBlue}
\cite{WangTWC} 2023 &
Multi-user/ single-target JCAS &
Sensing-at-STARS (sensor-embedded) &
Min CRB subject to comm QoS &
Space partition, beamforming, STARS coeffs &
AO, closed-form &
CRB/DoA; requires sensor integration assumptions \\
\hline

\rowcolor{BodyBlue}
\cite{11204524} 2025 &
Multi-BS multi-user/ multi-target JCAS &
Multi-BS cooperative JCAS with STAR-RIS &
Max--min sensing metric under comm QoS &
Active/passive beamforming across BSs &
AO, SDR &
Min sensing metric, QoS; coordination overhead issue \\
\hline

\rowcolor{BodyBlue}
\cite{xue2025stars} 2025 &
Single-user/ single-target near-field JCAS &
STARS-assisted near-field with sensors &
Min SPEB/CRB-type sensing error with QoS &
Sensor placement, beamforming, STARS coeffs &
AO (SCA/SDR-style) &
SPEB/CRB, QoS; highlights near-field modeling importance \\
\hline
\end{tabular}
\end{table*}

\subsection{Architectural Designs and Optimization Strategies in STAR-RIS Enabled JCAS}
This part looks over the newest developments that look at STAR-RIS structural changes and the new optimizations that go with them: convex programming, alternating optimization (AO), fractional programming (FP), and semidefinite relaxation (SDR), with an emphasis on the balancing trade-off in the sensing–communication and the design at the system level.

One fundamental architecture presented in \cite{ZehraHybrid} devised a hybrid STAR-RIS JCAS system that integrated passive reflective and active transmissive components to bypass multi-hop path attenuation and realize full-space electromagnetic coverage. This architecture differs from conventional approaches where sensing and communication are decoupled. With the help of this dual-sided STAR-RIS, integrated bidirectional communication and sensing are possible. A comprehensive mathematical derivation of the SIR and the corresponding SINR equations for both the sensing and communication components led to the formulation of a jointly non-convex optimization problem over the reflective and the transmissive coefficients. The problem was then formulated as a quadratically constrained quadratic program (QCQP) and was processed through Semidefinite Relaxation (SDR) using the CVX optimization toolbox. The performance of the system was evaluated through the Cramér--Rao bound (CRB) for the estimation of two-dimensional angle-of-departure (AoD), with numerical results showing notable gains over passive RIS benchmark models.
Similarly, to extend STAR-RIS capabilities to dual-function radar communication (DFRC), \cite{P-RISOJCOM} proposed a hybrid setup of STAR-RIS and passive-RIS (P-RIS) optimized through joint beamforming and phase-shift design. A weighted sum-rate maximization problem was constructed under power and sensing constraints and reformulated as a weighted minimum mean square error (WMMSE) problem. A low-complexity AO strategy enabled iterative updates, and the hybrid system achieved measurable improvements over conventional P-RIS and random phase-shift baselines, while still relying on ideal CSI.

Beyond hybrid STAR-RIS/P-RIS deployments, several recent works focused on \emph{beamforming-centric} STAR-RIS JCAS architectures that maximize sensing quality under communication QoS and hardware constraints. In particular, \cite{11059504} studied a STAR-RIS-assisted JCAS system and maximized the radar SINR subject to communication SINR constraints, a transmit power budget and STAR-RIS energy-splitting constraints, and developed an FP plus block coordinate descent (BCD) procedure (with SDR-based relaxations) to handle the resulting non-convex design. Complementarily, \cite{10978152} considered optimizing JCAS performance aided by STAR-RIS when sensing and communication share time/frequency/space resources, and further accounted for echo interference from communication users to the sensing target. The authors optimized active beamforming, STAR-RIS transmission/reflection coefficients, and multicarrier assignment variables under an energy-splitting protocol using an AO routine, and reported improved sensing performance relative to benchmark designs.

A second line of research strengthened architectural robustness by moving beyond deterministic perfect-CSI formulations and by targeting scalable large-antenna regimes. For example, \cite{10973322} investigated a STAR-RIS enabled JCAS setting under \emph{statistical CSI} and optimized BS beamforming and STAR-RIS phase control to characterize the ergodic-rate versus sensing-SCNR trade-off. To address channel uncertainty more explicitly, \cite{11099520} developed a \emph{robust} STAR-RIS-assisted MU-MIMO JCAS design and analyzed different operating protocols (e.g., time-splitting versus energy-splitting) while optimizing transceiver parameters to enhance JCAS performance under practical impairments. Moreover, to reduce computational burden in large arrays, \cite{10978628} studied a STAR-RIS-assisted massive MIMO JCAS system using hybrid partial zero-forcing (PZF) precoding, and derived closed-form expressions for communication sum-rate and a sensing metric (e.g., mean-to-average sidelobe ratio), highlighting the performance--complexity trade-off in large-scale implementations.

In addition to single-node optimization, recent architectures extended STAR-RIS-enabled JCAS toward \emph{multinode cooperative} sensing/communication. In \cite{WangVTC} and \cite{WangTWC}, a partitioned-space STAR-RIS JCAS framework was proposed that integrated a sensor-on-STARS structure with embedded sensors to mitigate clutter and improve detection. STAR-RIS coverage was divided into dedicated sensing and communication regions, and localization performance was optimized through CRB-oriented designs. Similarly, \cite{SongCBR} investigated CRB-driven space partitioning via joint beamforming and STAR-RIS control, minimizing the CRB while guaranteeing minimum SINR to communication users using AO with SDR and successive convex approximation (SCA). Extending cooperative sensing to multi-receiver settings, \cite{11174782} proposed a STAR-RIS-aided cooperative sensing framework that jointly optimized BS transmit beamforming, STAR-RIS transmission/reflection coefficients, and multi-receiver combining while also selecting sensing receivers; the resulting non-convex problem was handled via an AO/FP/SDR-based procedure with receiver-selection integration. Similarly, \cite{11204524} proposed a STAR-RIS-assisted multi-base-station (multi-BS) cooperative JCAS mechanism and jointly optimized active and passive beamforming to maximize the minimum sensing performance across BSs under communication QoS and power constraints, again relying on AO and SDR to address non-convex coupling across nodes.

Finally, several works explored STAR-RIS architectures under \emph{full-duplex and uplink} JCAS settings and \emph{near-field} propagation, both of which are important for future dense and high-frequency deployments. Specifically, \cite{11023100} introduced a full-duplex JCAS architecture enabled with STAR-RIS that aims to reduce self-interference to levels comparable to conventional cancelation methods and developed AO routines based on SDR/SCA to optimize transmit beamforming and STAR-RIS coefficients for sensing-SINR and sum-rate objectives. From an uplink perspective, \cite{11045175} studied STAR-RIS-assisted uplink JCAS with monostatic sensing under energy-splitting operation, and proposed a low-complexity beamforming approach (and extensions to multiple sensing-target/communication-user pairing via nonlinear optimization and matching). Regarding near-field designs, \cite{xue2025stars} proposed a STARS-assisted near-field JCAS framework where sensors are installed on the surface to enable distance-domain sensing with spherical wavefronts; the authors derived a squared position error bound (SPEB) expression and jointly optimized sensor deployment and beamforming under communication and power-consumption constraints. In addition, \cite{wang2025large} investigated a large-scale STAR-RIS-assisted near-field JCAS transmission method and optimized precoding/covariance and STAR-RIS transmission/reflection coefficients to minimize CRB-driven sensing error while maintaining communication QoS, thus highlighting the emerging importance of near-field modeling and scalable optimization in STAR-RIS-assisted JCAS.
\begin{table*}[!t]
\centering
\caption{Summary of waveform design work for STAR-RIS enabled JCAS.}
\label{tab:IV_B_summary}
\renewcommand{\arraystretch}{1.08}
\setlength{\tabcolsep}{5.2pt}


\begin{tabular}{|p{0.7cm}|p{2.7cm}|p{2cm}|p{2.7cm}|p{3.0cm}|p{1.5cm}|p{3cm}|}
\hline
\rowcolor{HeaderRed}
\textbf{Ref.} & \textbf{Waveform / Signal model} & \textbf{Scenario / Setup} & \textbf{Objective} & \textbf{Optimized Variables} & \textbf{Method} & \textbf{Notes / Metrics} \\
\hline

\rowcolor{BodyBlue}
\cite{Jifa-PIMRC} 2024 &
Learning-driven waveform design &
STAR-RIS JCAS (dynamic conditions) &
Reduce MUI while preserving waveform fidelity &
Waveform, STAR-RIS coefficients &
TD3 (MDP-based DRL) &
Sensing accuracy, comm. rate; scalability \\
\hline

\rowcolor{BodyBlue}
\cite{Jifa-TWC} 2024 &
Dual-functional constant-modulus waveform &
STAR-RIS JCAS (independent vs coupled phases) &
Min weighted sum of MUI and waveform deviation &
Waveform, BS beamforming, STAR-RIS coefficients &
ADMM (independent) / TD3 (coupled) &
Highlights trade-off control; DRL generalization remains open \\
\hline

\rowcolor{BodyBlue}
\cite{taremizadeh2025star} 2025 &
Pulsed-signal transceiver/waveform co-design &
STAR-RIS JCAS with pulsed signaling &
Improve sensing performance under comm QoS &
Pulse waveform parameters, STAR-RIS coefficients (and possibly beamforming) &
AO / convexification (problem-dependent) &
Radar-aligned signaling; supports temporal separation/pulse compression \\
\hline

\rowcolor{BodyBlue}
\cite{11242204} 2025 &
Space--time coded pulsed signals &
STAR-RIS JCAS with structured pulsed waveform &
Enhance sensing diversity while maintaining comm reliability &
Pulse/code structure, STAR-RIS coefficients (and transceiver parameters) &
Optimization-based iterative design &
Improves echo-domain structure; added coding/implementation complexity \\
\hline

\rowcolor{BodyBlue}
\cite{11129028} 2025 &
DFRC waveform/precoder trade-off design &
STAR-RIS-aided DFRC &
Balance sensing (beampattern/detection) and communication QoS &
Transmit strategy/precoder, STAR-RIS configuration &
Optimization-based (AO/SDR/SCA style) &
Tunable sensing--comm operating point; complements DRL-based designs \\
\hline

\end{tabular}
\end{table*}

\noindent\textbf{Discussion and insights:}
The reviewed works show a progression from foundational STAR-RIS architectures to more realistic and scalable designs. Early frameworks exploit full-space coverage and joint beamforming/STAR-RIS configuration via SDR-based convexification (e.g., hybrid STAR-RIS and STAR-RIS+P-RIS deployments), demonstrating that programmable transmission/reflection can enlarge the sensing--communication feasible region. Beamforming-centric studies then introduce explicit trade-off objectives (e.g., radar SINR under communication SINR constraints) and rely on AO/FP/BCD-type procedures to address the strong coupling between active and passive variables. More recent efforts relax ideal-CSI assumptions through statistical/robust CSI and propose lower-complexity structures for large arrays (e.g., PZF-type precoding), underscoring the need to balance performance gains with solver complexity and signaling overhead. A further trend is the shift from single-node optimization to cooperative and deployment-aware architectures, including sensor-embedded/partitioned-space STARS for clutter mitigation and CRB-driven sensing, as well as multi-BS cooperation to improve sensing reliability under full-spectrum reuse. Near-field modeling and sensor deployment also emerge as key directions for dense, high-frequency systems where spherical wavefronts affect both sensing and the optimal STAR-RIS configuration. These trends are summarized in Table~\ref{tab:IV_A_summary}, which also highlights recurring issues such as imperfect CSI, coefficient-coupling constraints, coordination overhead, and the need for scalable real-time optimization.

\subsection{Waveform Design for STAR-RIS Enabled JCAS}
Complementing system-level beamforming and coefficient optimization, waveform design has emerged as a critical component in STAR-RIS-aided JCAS, since the transmitted signal structure directly determines both multi-user interference (MUI) behavior and sensing fidelity.

\subsubsection{Learning-driven waveform design under dynamic conditions}
In \cite{Jifa-PIMRC}, the authors employ the twin delayed deep deterministic policy gradient (TD3) algorithm within a Markov decision process (MDP) framework to jointly optimize STAR-RIS coefficients and waveform design under dynamic network conditions. Your comments about balancing real-time adaptability and keeping waveforms unaltered to reduce MUI are appreciated. Compared to other learning-based JCAS methods, this model-free reinforcement learning approach shows the most promise when it comes to sensing and communicating. Yet, it raises the worry of computational scalability and convergence, especially when faced with limited or faulty channel state information (CSI) and hardware. Building on this, \cite{Jifa-TWC} constructs a dual-purpose constant-modulus waveform for STAR-RIS-aided JCAS that balances communication and sensing. The authors constructs a linear unconstrained model to balance MUI and waveform deviation. The ADMM is applied to the independent phase-shift model, and for the more complex coupled model. The authors also investigate reinforcement learning based optimation and study the multi-objective compared to conventional RIS. It is empirically verified that STAR-RIS offers greater capacity of managing multiple objectives than traditional RIS.

\subsubsection{Transceiver/waveform co-design with pulsed signaling}
Beyond continuous-waveform formulations, recent works have investigated \emph{pulsed} and structured waveforms that are more radar-aligned while remaining communication compatible. In \cite{taremizadeh2025star}, STAR-RIS transceivers are designed for JCAS using pulsed signals, where waveform parameters and STAR-RIS coefficients are co-optimized to improve sensing performance while preserving communication QoS. This line of work is particularly relevant for scenarios where sensing requires temporal separation or pulse compression gains, and it helps bridge the gap between radar-style signaling and STAR-RIS-enabled communication links. Moreover, \cite{11242204} further enhances pulsed STAR-RIS JCAS by introducing space--time coded pulsed signals, enabling additional diversity/structure in the echo domain while maintaining communication reliability. Such designs provide an alternative to purely power-/phase-only optimization by explicitly shaping the transmitted waveform to control both sensing ambiguity and inter-user interference.

\subsubsection{DFRC-oriented waveform balancing and trade-off control}
From a DFRC perspective, balancing the communication and sensing objectives has been done using STAR-RIS-aided DFRC waveform/precoder optimization in \cite{11129028}. The solution provides tunable configurations in communications throughput/quality and in sensing performance (e.g., beampattern and/or detection metrics) by jointly optimizing the scaled aperture reflectors (STAR) transmits and the range (RIS) reflect) adjuster configurations. This DFRC perspective balances the DRL-based constant modulus waveform designs and provides the sense/communicate trade-off optimization mechanisms while also noting that STAR-RIS positively alters the sensing-communicating trade-off compared to usual RIS.

\noindent\textbf{Discussion and insights:}
The reviewed works indicate that waveform design in STAR-RIS-aided JCAS is evolving from purely coefficient/beamforming optimization toward explicitly shaping the transmitted signal to control both sensing fidelity and multi-user interference. A first trend is the growing use of learning-based waveform adaptation (e.g., TD3-driven designs) to cope with time-varying environments and coupled STAR-RIS constraints; while these methods can improve performance in dynamic settings, their convergence, generalization, and real-time feasibility under imperfect CSI remain critical concerns. A second trend is the re-emergence of radar-aligned signaling, especially pulsed and space--time coded pulsed waveforms, which naturally support sensing tasks (e.g., temporal separation and echo-domain structure) while maintaining communication QoS, albeit at the cost of additional signaling/implementation complexity. Finally, DFRC-oriented formulations provide an optimization-driven mechanism to tune the sensing--communication operating point (e.g., beampattern/detection versus rate), and they serve as a complementary alternative to DRL approaches when model-based optimization is feasible. These representative waveform-design directions, along with their objectives, optimized variables, and solution methodologies, are summarized in Table~\ref{tab:IV_B_summary}, which highlights a key open challenge: achieving robust, low-complexity waveform and STAR-RIS configuration updates that remain effective under hardware coupling, imperfect CSI, and stringent latency constraints.

\subsection{Security and Privacy in STAR-RIS Enabled JCAS}
Security and privacy in STAR-RIS-aided JCAS have attracted strong interest because simultaneous transmission and reflection enlarge the attack surface (eavesdropping, detection, and jamming), while sensing constraints restrict how aggressively communication can be protected. Recent works therefore design STAR-RIS coefficients jointly with transmit/receive processing to enhance secrecy and/or covertness without sacrificing sensing quality.
\begin{table*}[!t]
\centering
\caption{Summary of Security and privacy in STAR-RIS enabled JCAS works.}
\label{tab:IV_D_summary_part1}
\renewcommand{\arraystretch}{1.08}
\setlength{\tabcolsep}{4.5pt}


\begin{tabular}{|p{0.8cm}|p{3cm}|p{3cm}|p{3.2cm}|p{2.8cm}|p{3.5cm}|}
\hline
\rowcolor{HeaderRed}
\textbf{Ref.} & \textbf{Threat model} & \textbf{Scenario / Architecture} & \textbf{Objective} & \textbf{Method} & \textbf{Metrics / Notes} \\
\hline

\rowcolor{BodyBlue}
\cite{LIU-TVT} 2025 &
Covert detection (Willie), imperfect CSI &
STAR-RIS JCAS mmWave &
Max covert rate s.t.\ PDE and sensing constraints &
SCA/SDR-type iterative optimization &
PDE, covert rate; leverages channel uncertainty \\
\hline

\rowcolor{BodyBlue}
\cite{LIU-JIOT} 2025 &
Joint eavesdropping + covert detection &
\emph{Active} STAR-RIS JCAS &
Max secure sum-rate under PDE, QoS, sensing SINR &
SCA + SDR + SROCR &
Sum-rate, leakage/PDE; higher complexity \\
\hline

\rowcolor{BodyBlue}
\cite{10974475} 2025 &
Joint secure \& covert constraints &
Active STAR-RIS JCAS &
Joint secrecy/covertness optimization under sensing QoS &
Optimization-based iterative design (AO/SCA style) &
Secrecy + covert constraints; highlights feasible-region gain \\
\hline

\rowcolor{BodyBlue}
\cite{zhang2025star} 2025 &
Covert constraint + computation privacy risk &
STAR-RIS JCAS + AirComp (ISCCO) &
Balance covert comm/sensing and AirComp aggregation &
Analysis + optimization framework &
Low probability of detection + aggregation performance trade-off \\
\hline

\rowcolor{BodyBlue}
\cite{Wei-ICC} 2024 &
Eavesdropping risk, sensing beampattern constraint &
STAR-RIS JCAS (region partition, NOMA-enabled) &
Max sum secrecy rate with sensing beampattern constraints &
AO + SCA (convexified subproblems) &
Secrecy rate; AN/jamming may impact sensing \\
\hline

\rowcolor{BodyBlue}
\cite{Wei-TWC} 2024 &
Eavesdropping at sensing target; AN/jamming  &
STAR-RIS JCAS (partitioned space) &
Joint secrecy enhancement and sensing constraints &
AO + SCA &
Secrecy rate + sensing QoS; interference coupling \\
\hline

\rowcolor{BodyBlue}
\cite{Chao-ICC} 2023 &
Malicious radar target / eavesdropper &
Secure STAR-RIS-aided DFRC &
Max radar sensing power with secure comm constraints &
Symbol-level precoding + iterative solver &
Security via constructive/destructive interference control \\
\hline

\rowcolor{BodyBlue}
\cite{Chao-TIFS} 2023 &
Secure DFRC under interference exploitation &
STAR-RIS-aided DFRC &
Secure comm + radar objective balancing &
Distance-majorization iterative algorithm &
Implementation complexity \\
\hline

\rowcolor{BodyBlue}
\cite{TAO-JIOT} 2025 &
Imperfect CSI + eavesdropping, IoE security &
STAR-RIS JCAS (CE vs CSF; ES/MS; coupled/indep.) &
Robust secure JCAS with AN-aided transmission &
Penalty/BCD + worst-case robust design &
QoS + secrecy; protocol-dependent performance \\
\hline

\rowcolor{BodyBlue}
\cite{11144762} 2025 &
Dual eavesdropping attacks under coupling constraints &
Coupled phase-shift STAR-RIS JCAS &
Max secrecy rate with coupled STAR-RIS constraints &
Iterative optimization (AO/SCA/SDR style) &
Explicitly models multiple adversaries + coupling \\
\hline

\rowcolor{BodyBlue}
\cite{kazymova2025achievable} 2025 &
Secrecy in multi-user JCAS (MU-MIMO) &
STAR-RIS-enabled MU-MIMO JCAS &
Analyze/optimize sum secrecy rate &
Optimization + analysis (model-based) &
Secrecy rate; highlights spatial multiplexing effects \\
\hline

\rowcolor{BodyBlue}
\cite{Xie-WCSP} 2024  &
Jammer (modeled as extended sensing target) &
STAR-RIS JCAS under jamming &
Max rate with CRB and QoS constraints &
AO (FP) + SDR/Taylor &
Rate + CRB; jammer modeling realism noted \\
\hline

\rowcolor{BodyBlue}
\cite{Lin-SCIC} 2024 &
Interception risk; secure scheduling trade-off &
STAR-FC-RIS (fully connected) JCAS &
Scheduling + time allocation for security/reliability &
Closed-form outage/ intercept analysis, OTA &
Sensing outage, comm outage, intercept probability \\
\hline

\rowcolor{BodyBlue}
\cite{11079638} 2025 &
Covert JCAS with deployment uncertainty &
Movable-element STAR-RIS JCAS &
Robust covert design via joint beamforming &
Robust optimization (AO/SCA/SDR style) &
Covertness metric + QoS; adds control overhead \\
\hline

\rowcolor{BodyBlue}
\cite{chen2025secrecy} 2025 &
Secrecy enhancement via hardware mobility &
Movable-antenna-aided STAR-RIS JCAS &
Secrecy rate via antenna positioning, STAR-RIS design &
Joint optimization (iterative) &
Secrecy rate; mechanical feasibility considerations \\
\hline

\rowcolor{BodyBlue}
\cite{11091501} 2026 &
Passive eavesdropper with sensing-assisted refinement &
JCAS-aided secure comm with STAR-RIS &
Counter ``self-refine sensing'' eavesdropping &
Optimization-based secure design &
Addresses stronger adversary model beyond passive listening \\
\hline

\rowcolor{BodyBlue}
\cite{noh2025multiple} 2025 &
Eavesdropping in multi-active deployments &
Multi-active STAR-RIS secure JCAS &
Enhance secrecy under sensing constraints &
Cooperative/robust beamforming optimization &
Secrecy + sensing QoS; cooperation benefits \\
\hline

\rowcolor{BodyBlue}
\cite{11189492} 2025 &
Secure JCAS via cooperative STAR-RIS design &
STAR-RIS secure JCAS (cooperation) &
Improve secrecy while preserving sensing requirements &
Cooperative/robust beamforming &
Secrecy performance; coordination overhead remains \\
\hline

\rowcolor{BodyBlue}
\cite{Zhu-TST} 2025 &
Long-term secrecy under time-varying channels &
STAR-RIS JCAS (ES/TS, coupled phases) &
Max long-term secrecy rate  &
DRL (DDPG / SAC) &
Secrecy rate; training cost vs real-time solvability \\
\hline

\end{tabular}
\end{table*}

\subsubsection{Covert JCAS and joint secure-covert designs}
In \cite{LIU-TVT}, the authors studied covert communication in STAR-RIS-aided JCAS mmWave networks under imperfect CSI, derived detection error probability (PDE) expressions for an adversary, and optimized active/passive beamforming while satisfying sensing and covertness constraints via SCA/SDR-type iterations. In \cite{LIU-JIOT}, the same line of work was extended to an \emph{active} STAR-RIS network facing simultaneous detection and eavesdropping threats, where secure and covert transmissions were jointly supported under SINR and QoS detection constraints using iterative SCA/SDR/SROCR updates. Complementing these, \cite{10974475} investigated joint secure and covert communications in active STAR-RIS.assisted JCAS by jointly optimizing beamforming and STAR-RIS coefficients under covertness and secrecy requirements, illustrating how STAR-RIS can expand the feasible region compared to conventional RIS. Moreover, \cite{zhang2025star} integrated covertness with sensing/communication and air computing (AirComp), proposing a STAR-RIS-empowered covert JCAS-computation framework with analysis and optimization to balance computing aggregation performance and low probability of detection.

\subsubsection{Secrecy beamforming and robustness against advanced eavesdroppers}
Beyond covert transmission, several works focus on improving secrecy rates under sensing constraints and stronger attack models. In \cite{Wei-ICC,Wei-TWC}, the authors formulated the maximization of the sum secrecy rate for JCAS assisted by STAR-RIS by jointly optimizing the BS beamforming, artificial jamming, and STAR-RIS transmission/reflection coefficients under sensing beampattern constraints, and solved the resulting non-convex problem via AO/SCA-based convexification. From a different physical-layer angle, \cite{Chao-ICC,Chao-TIFS} proposed symbol-level precoding for secure STAR-RIS-aided DFRC, exploiting constructive/destructive interference to enhance legitimate reception while misleading malicious targets, and developed low-complexity iterative solvers. In IoE-oriented settings, \cite{TAO-JIOT} considered robust secure STAR-RIS-assisted JCAS under imperfect CSI and artificial-noise (AN) injection for different STAR-RIS protocols, again using penalty/BCD-type iterations to maintain QoS while suppressing information leakage. In addition, \cite{11144762} addressed \emph{dual eavesdropping attacks} under coupled STAR-RIS phase-shift constraints and optimized secrecy performance in the presence of multiple adversaries, highlighting the need for robust designs when attacker capabilities and coupling constraints coexist. For MU-MIMO, \cite{kazymova2025achievable} analyzed/optimized the achievable sum secrecy rate of the STAR-RIS-enabled MU-MIMO JCAS, offering insight into how spatial multiplexing and STAR-RIS coefficients interact under secrecy constraints.

\subsubsection{Anti-jamming and secure scheduling in JCAS}
Security in STAR-RIS JCAS is also challenged by intentional jamming, which directly degrades both sensing and communications. In \cite{Xie-WCSP}, a STAR-RIS-aided JCAS framework was studied in the presence of a jammer, modeled as an extended sensing target, and joint active/passive beamforming was designed to meet CRB and user-rate constraints using AO with FP/SDR-type updates. Although effective numerically, the model motivates future work on more realistic adaptive jamming behaviors and lower-complexity anti-jamming solutions. Furthermore, \cite{Lin-SCIC} investigated secure scheduling for a fully connected STAR-RIS (STAR-FC-RIS)- aided JCAS network and proposed multiuser scheduling and time-allocation strategies to balance sensing reliability, communication outage, and interception probability via closed-form outage/intercept analysis.

\subsubsection{Hardware mobility - movable elements/antennas and sensing-assisted eavesdropper models}
A recent trend is to enhance secrecy/covertness by extending STAR-RIS control beyond phase design to \emph{hardware mobility}. In particular, \cite{11079638} introduced movable STAR-RIS elements for covert JCAS and jointly optimized robust beamforming and element deployment, improving covertness under uncertainty. Similarly, \cite{chen2025secrecy} studied STAR-RIS JCAS aided by moving antennas and maximized the secrecy rate by jointly optimizing the antenna positions and the STAR-RIS coefficients. Additionally, \cite{11091501} explored a JCAS-aided secure communication model where a passive eavesdropper refines its attack using sensing (``self-refine sensing''), and leveraged STAR-RIS-assisted design to counter this stronger adversary. Finally, \cite{noh2025multiple} and \cite{11189492} studied secure STAR-RIS JCAS via cooperative/robust beamforming, including multi-active STAR-RIS deployments and cooperative beamforming strategies, showing how spatial cooperation can improve secrecy while maintaining sensing constraints.

\subsubsection{Learning-based long-term secure optimization}
Finally, in \cite{Zhu-TST} the authors proposed a dual-secure STAR-RIS-assisted JCAS framework and used DRL (e.g., DDPG/SAC) to optimize long-term secrecy performance under time-varying channels and coupled phase-shift constraints, providing a practical alternative when repeated convex optimization is computationally heavy.

\noindent\textbf{Discussion and insights:}
The security and privacy literature on STAR-RIS-aided JCAS exhibits several clear trends. First, STAR-RIS enables covert JCAS designs where covertness (e.g., low detection probability) must be met jointly with sensing SINR and communication QoS, yielding tightly coupled non-convex problems typically handled by iterative convexification (SCA/SDR) or, for long-term operation, DRL-based policies. Second, secrecy-focused studies increasingly adopt stronger adversary models (e.g., dual/passive eavesdroppers, sensing-aided eavesdropping, and active jamming), highlighting that sensing can be both a vulnerability and a defense depending on how transmit/reflect components are shaped. Third, designs are expanding beyond phase/beamforming control to additional knobs such as scheduling/time allocation in fully connected STAR-RIS architectures and hardware mobility (movable elements/antennas), improving the secrecy/covertness region at the cost of higher control overhead and implementation complexity. Table~\ref{tab:IV_D_summary_part1} summarizes the main contributions. Most existing methods assume accurate CSI (or bounded uncertainty) and rely on computationally intensive iterations, which may be challenging under short coherence times. Moreover, sensing--security coupling creates inherent trade-offs: AN/jamming can enhance secrecy but degrade sensing, while stricter covertness often reduces throughput. Future work should prioritize (i) robust secure/covert designs under imperfect/outdated CSI and hardware coupling, (ii) low-complexity scalable optimization/learning with reliability guarantees, and (iii) realistic adaptive attacker models and validation under practical update and signaling constraints.

\begin{table*}[!t]
\centering
\caption{Energy efficiency (EE) and sustainability-oriented works in STAR-RIS enabled JCAS (Section IV-F).}
\label{tab:IV_F_EE_summary}
\footnotesize
\renewcommand{\arraystretch}{1.08}
\setlength{\tabcolsep}{4.8pt}


\begin{tabular}{|p{0.7cm}|p{2.5cm}|p{2.5cm}|p{2.4cm}|p{2.8cm}|p{2cm}|p{3cm}|}
\hline
\rowcolor{HeaderRed}
\textbf{Ref.} & \textbf{Scenario / Setup} & \textbf{Architecture / Protocol} & \textbf{Objective} & \textbf{Optimized Variables} & \textbf{Method} & \textbf{Metrics / Notes} \\
\hline

\rowcolor{BodyBlue}
\cite{QIN-CL} 2025 &
STAR-RIS JCAS (deployment-aware) &
STAR-RIS with location optimization &
Max EE under sensing/comm constraints &
BS beamforming, STAR-RIS T/R coeffs, placement/location &
AO, SDR, fractional programming &
EE vs sensing threshold trade-off; placement is critical \\
\hline

\rowcolor{BodyBlue}
\cite{11134414} 2025 &
Near-field, low-altitude JCAS &
Near-field STAR-RIS-enhanced low-altitude network &
Energy-efficient beamforming with QoS &
Beamforming, STAR-RIS coeffs &
AO, SCA, SDR &
Near-field modeling shifts EE-optimal operating point \\
\hline

\rowcolor{BodyBlue}
\cite{11150384} 2025 &
Secure EE under uncertainty &
\emph{Active} STAR-RIS JCAS (robust) &
Max secure EE with sensing, secrecy constraints &
Beamforming, active STAR-RIS variables (and robust margins) &
AO, SCA, SDR &
Highlights circuit-power impact; secrecy can reduce EE \\
\hline

\rowcolor{BodyBlue}
\cite{10891707} 2025 &
MIMO JCAS with energy harvesting &
Active, STAR-RIS-assisted MIMO JCAS with SWIPT &
Joint sensing--comm--EH design &
Transmit strategy, STAR-RIS config (+ EH constraints) &
Iterative design &
Trade-off among rate, sensing, and harvested energy \\
\hline

\rowcolor{BodyBlue}
\cite{kumar2025green} 2025 &
Green JCAS with power transfer &
STAR-RIS-enabled JCAS, power transfer &
Integrate sensing, communication, and power transfer &
STAR-RIS T/R control, system resource variables &
Analysis, optimization framework &
Sustainability beyond EE; emphasizes power-transfer trade-offs \\
\hline

\end{tabular}
\end{table*}

\subsection{Energy Efficiency in STAR-RIS Enabled JCAS}
Energy efficiency (EE) has become a critical design objective in STAR-RIS-aided JCAS systems, particularly for sustainable and power-constrained 6G deployments. By enabling full-space transmission and reflection, STAR-RIS introduces additional degrees of freedom for energy-aware beamforming and resource allocation but also creates new trade-offs among EE, sensing quality, hardware power consumption, and signaling overhead. This subsection reviews recent EE-centric work in STAR-RIS-enabled JCAS, highlighting how EE objectives are formulated and optimized under sensing and communication constraints.

\subsubsection{EE Maximization and Optimal STAR-RIS Placement} In \cite{QIN-CL}, a representative EE optimization framework is proposed, where EE is improved by jointly optimizing the BS transmit beamforming, the STAR-RIS transmission/reflection coefficients, and the location of the STAR-RIS deployment. The resulting non-convex problem is handled via AO by decomposing it into subproblems solved iteratively using SDR and fractional programming. Numerical results demonstrate that STAR-RIS can significantly improve EE over conventional RIS configurations and that location optimization plays an important role. The study also reveals a fundamental trade-off: enforcing stricter sensing (e.g., higher beampattern gain requirements) improves sensing accuracy but reduces EE.

\subsubsection{Near-field and low-altitude EE design}
Beyond far-field modeling, \cite{11134414} investigated energy-efficient beamforming for near-field STAR-RIS-enhanced low-altitude JCAS networks, where spherical wavefront effects and short-range geometry become important. The work optimizes active beamforming and STAR-RIS coefficients under sensing/communication QoS and power constraints, showing that near-field modeling can change the optimal EE operating point and that low-altitude STAR-RIS deployment can provide EE gains when designed jointly with beamforming.

\subsubsection{Secure EE under active STAR-RIS}
Although most EE works focus on power consumption and QoS, security constraints can further tighten the feasible design space and reduce EE if not handled carefully. In \cite{11150384}, the authors studied robust secure EE optimization for \emph{active} STAR-RIS-assisted JCAS, jointly considering EE maximization, sensing requirements, and secrecy/robustness under uncertainty. Such formulations highlight that active STAR-RIS offers additional control flexibility, but may also incur extra circuit power, making accurate power modeling essential for meaningful EE conclusions.

\subsubsection{Power transfer and SWIPT-enabled sustainability}
A growing direction is to integrate energy transfer into the STAR-RIS JCAS, linking sustainability not only to reduced consumption but also to \emph{wireless power delivery}. In \cite{10891707}, active and STAR-RIS-assisted MIMO JCAS with SWIPT is investigated, where the design jointly supports the sensing, communication and energy harvesting objectives. Similarly, \cite{kumar2025green} proposed a STAR-RIS-based framework for green integration of sensing, communication, and power transfer, illustrating how STAR-RIS can coordinate reflection/transmission to simultaneously serve information users, sensing targets, and energy harvesting nodes. These works reveal additional trade-offs among harvested energy, sensing quality, and communication rate, and motivate future research on practical power-consumption models (including STAR-RIS controller power), real-time adaptation, and imperfect CSI.

\noindent\textbf{Discussion and insights:}
The EE-centric STAR-RIS JCAS literature increasingly treats sustainability as a primary objective rather than a by-product of link-budget gains. As summarized in Table~\ref{tab:IV_F_EE_summary}, early works maximize bits/Joule under sensing and communication constraints and expose a fundamental trade-off: tighter sensing requirements (e.g., higher beampattern gain or sensing-SINR thresholds) typically lower EE due to higher transmit power and/or more restrictive beam patterns. Another trend concerns deployment and propagation regimes: near-field and low-altitude scenarios alter the effective channel structure and can shift the EE-optimal operating point, motivating geometry-aware design and placement optimization. Incorporating security and robustness further highlights that imperfect CSI and secrecy constraints can tighten feasibility and reduce EE unless properly balanced. Finally, SWIPT and power-transfer-enabled STAR-RIS JCAS extend sustainability beyond energy savings to energy delivery, creating new trade-offs among harvested energy, sensing quality, and communication throughput. Overall, Table~\ref{tab:IV_F_EE_summary} suggests prioritizing holistic power-consumption models (including STAR-RIS controller/circuit power), low-overhead online adaptation, and robust EE formulations that remain effective under CSI uncertainty and practical hardware constraints.

\subsection{NOMA and RSMA-based STAR-RIS Enabled JCAS}
Integrating non-orthogonal multiple access (NOMA) with STAR-RIS-enabled JCAS improves spectral efficiency and user connectivity by exploiting full-space transmission/reflection while enabling simultaneous service of communication users (CUs) and sensing users/targets (SUs). However, the joint design introduces additional challenges, including strong interference coupling (between sensing and communication signals), fairness constraints, SIC ordering, hardware-induced coupling of the STAR-RIS coefficient, and computational complexity. This subsection reviews recent NOMA-/RSMA-based STAR-RIS JCAS studies, progressing from RSMA-based formulations to fairness-oriented NOMA designs and scalable clustering-/learning-based solutions.

\subsubsection{RSMA-based STAR-RIS JCAS} 
The authors in \cite{LIU-XIV} explored STAR-RIS-enabled JCAS with multiple access rate splitting (RSMA) under an energy-splitting protocol, formulating a sensing-SINR maximization problem by joint optimization of RSMA rate splitting, BS beamforming, and STAR-RIS coefficients. The resulting non-convex problem was addressed through a decomposition pipeline involving SDR-type relaxations and iterative updates. Complementing this direction, \cite{10949621} further studied RSMA-assisted STAR-RIS JCAS and jointly optimized rate-splitting variables and active/passive beamforming to better manage multi-user interference under sensing constraints. From an analytical perspective, \cite{10964255} provided performance analysis for MISO-JCAS aided by STAR-IRS with multiple targets under a rate-splitting approach, offering useful information on achievable trade-offs and design trends in multi-target scenarios. In addition, \cite{to2025fairness} integrated RSMA with a secure design with fairness in STAR-RIS JCAS, highlighting the interplay between secrecy, fairness and sensing constraints.
\begin{table*}[!t]
\centering
\caption{Summary of NOMA/RSMA-based STAR-RIS enabled JCAS works.}
\label{tab:IV_E_summary_all}
\footnotesize
\renewcommand{\arraystretch}{1.08}
\setlength{\tabcolsep}{4.6pt}


\begin{tabular}{|p{0.7cm}|p{1cm}|p{2.55cm}|p{3.0cm}|p{2.65cm}|p{2.5cm}|p{3.25cm}|}
\hline
\rowcolor{HeaderRed}
\textbf{Ref.} & \textbf{MA} & \textbf{Scenario / Setup} & \textbf{Objective} & \textbf{Optimized Variables} & \textbf{Method} & \textbf{Metrics / Notes} \\
\hline

\rowcolor{BodyBlue}
\cite{LIU-XIV} 2024 &
RSMA &
STAR-RIS JCAS (ES mode) &
Max sensing SINR under comm constraints &
Rate-splitting + BS beamforming + STAR-RIS coeffs &
SDR + Dinkelbach + SROCR (iterative) &
Sensing SINR; highlights RSMA interference management \\
\hline

\rowcolor{BodyBlue}
\cite{10949621} 2026 &
RSMA &
STAR-RIS JCAS (MU setting) &
Joint RSMA/beamforming under sensing constraints &
Rate-splitting + active/passive beamforming &
Iterative optimization (AO/SDR/SCA-style) &
Trade-off control; coupling/complexity concerns \\
\hline

\rowcolor{BodyBlue}
\cite{10964255} 2025 &
RSMA &
STAR-IRS-aided MISO-JCAS (multi-target) &
Performance analysis of RSMA under sensing constraints &
(Analytical focus) beamforming/RS structure &
Analysis + derived expressions &
Rate--sensing trade-offs; multi-target insights \\
\hline

\rowcolor{BodyBlue}
\cite{to2025fairness} 2025 &
RSMA &
STAR-RIS JCAS with fairness/security &
Fairness-aware secure RSMA under sensing constraints &
Beamforming + RS variables (+ security terms) &
Iterative secure design (AO/SCA style) &
Fairness + secrecy + sensing interplay \\
\hline

\rowcolor{BodyBlue}
\cite{Yufei-TWC} 2024 &
NOMA &
STAR-RIS JCAS (fairness, SINR/SCNR) &
Max--min fairness across CUs and sensing users &
BS beamforming + STAR-RIS coeffs (+ SIC order) &
AO + SCA/SDP (closed-form in special case) &
Improved fairness; CSI/coupling assumptions noted \\
\hline

\rowcolor{BodyBlue}
\cite{Ruidong-VTC} 2024 &
NOMA &
STAR-RIS JCAS (fairness, CRLB-based) &
Balance CU SINR and sensing accuracy (CRLB) &
Beamforming + STAR-RIS coeffs (+ SIC-related decisions) &
Low-complexity AO + SCA/SDP &
CRLB + rate/fairness; relies on CSI tracking \\
\hline

\rowcolor{BodyBlue}
\cite{YU-TVT} 2024 &
NOMA &
STAR-NOMA-JCAS (multi-target, multi-user) &
Max average weighted sum rate (AWSR) &
User pairing/scheduling + active/passive beamforming &
Penalty/CCP-style decomposition &
AWSR; mixed discrete-continuous optimization \\
\hline

\rowcolor{BodyBlue}
\cite{XUE-GC} 2023 &
CB-NOMA &
STAR-RIS JCAS (cluster-based) &
Max min beampattern gain under comm constraints &
Power allocation + active/passive beamforming &
BCD + SCA + penalty method &
Beampattern gain, mismatch error; inter-cluster issues remain \\
\hline

\rowcolor{BodyBlue}
\cite{XU-TWC} 2024 &
CB-NOMA &
STAR-RIS JCAS (scalable clustering) &
Improve sensing/comm with reduced complexity &
Power + active/passive beamforming (element-wise update) &
BCD-M / BCD-E (SDR+SCA; element-wise updates) &
Near sense-only beampattern; lower complexity for large $M$ \\
\hline

\rowcolor{BodyBlue}
\cite{11202653} 2025 &
NOMA (DRL) &
Multi-active STAR-RIS NOMA-JCAS &
Joint design with discrete element selection &
Element selection + beamforming + STAR-RIS coeffs &
DRL (mixed discrete/continuous control) &
Adaptation potential; training/robustness issues \\
\hline

\rowcolor{BodyBlue}
\cite{11049876} 2025 &
SCMA (meta-DRL) &
STAR-IRS-aided SCMA-JCAS &
Joint beamforming and resource allocation &
Beamforming + resource allocation variables &
Meta deep reinforcement learning &
Fast adaptation across environments; overhead concerns \\
\hline

\rowcolor{BodyBlue}
\cite{11162314} 2025 &
NOMA (baseline) &
Multi-sector RIS-aided JCAS (not STAR-RIS) &
Analyze NOMA+JCAS under sectorization &
(Analysis/optimization depending on setup) &
Model-based analysis/optimization &
Useful baseline for sectorization vs STAR-RIS full-space \\
\hline
\end{tabular}
\end{table*}

\subsubsection{Fairness-aware NOMA-JCAS with STAR-RIS}
To meet fairness requirements in multi-user JCAS, \cite{Yufei-TWC} proposed a max--min framework that balances CU SINR and SU sensing quality (SCNR), explicitly accounting for the interference from sensing signals to communication links. The joint optimization of the BS beamforming and STAR-RIS coefficients was handled via AO (with SCA/SDP refinements), and a closed-form solution was derived for special cases. Similarly, \cite{Ruidong-VTC} introduced a CRLB-based fairness formulation, optimizing beamforming, STAR-RIS coefficients, and (implicitly) SIC-related decisions to trade off communication reliability and sensing accuracy. These works highlight that fairness metrics (SCNR/SINR/CRLB) provide a systematic way to quantify JCAS balance, while also exposing practical issues such as CSI tracking and STAR-RIS coefficient coupling.

\subsubsection{User pairing/scheduling and scalable multi-user designs}
In \cite{YU-TVT}, the authors introduced a STAR-NOMA-JCAS framework with multi-target sensing and multi-user communication, and maximized the average weighted sum rate by jointly optimizing user-pair scheduling, active beamforming and STAR-RIS coefficients under sensing constraints. The mixed discrete-continuous nature of pairing and rank constraints was addressed via penalty-style/CCP-style decomposition. These scheduling-based formulations are important for practical deployments where the user set changes over time, and sensing/communication priorities must be dynamically balanced.

\subsubsection{Cluster-based NOMA and complexity reduction}
To scale NOMA-JCAS to full-space coverage, \cite{XUE-GC} proposed a cluster-based NOMA (CB-NOMA) STAR-RIS JCAS design and optimized power allocation and active/passive beam formation to maximize minimum beam pattern gain under communication constraints. \cite{XU-TWC} further refined the CB-NOMA STAR-RIS JCAS by developing BCD-based algorithms, including an element-wise update strategy that reduces the complexity of passive beamforming optimization while maintaining sensing performance. These works demonstrate that clustering and low-complexity updates can make STAR-RIS-aided NOMA-JCAS more scalable, although inter-cluster interference and SIC robustness remain key bottlenecks.

\subsubsection{Learning-assisted NOMA/SCMA STAR-RIS JCAS}
Recent studies have also adopted learning to cope with the strong coupling and large design space in STAR-RIS JCAS. In \cite{11202653}, a multi-active STAR-RIS-aided NOMA-JCAS system incorporated element selection and used deep reinforcement learning (DRL) to jointly handle discrete (element selection) and continuous (beamforming/coefficient) actions. Moreover, \cite{11049876} investigated SCMA-JCAS assisted with STAR-IRS and employed meta deep reinforcement learning to jointly optimize beamforming and resource allocation, with the aim of improving adaptability across environments. Finally, \cite{11162314} analyzed NOMA-enabled multi-sector RIS-aided JCAS (not restricted to STAR-RIS) and can serve as a useful related baseline highlighting how sectorization and non-orthogonal access interact under sensing requirements.

Overall, the literature indicates that RSMA and NOMA provide complementary tools for managing interference and connectivity in STAR-RIS JCAS, while fairness-aware formulations (SINR/SCNR/CRLB) and scalable clustering/learning methods are key enablers for practical multi-user and multi-target operation.

\noindent\textbf{Discussion and insights:}
RSMA/NOMA-based STAR-RIS JCAS studies show that non-orthogonal access effectively exploits full-space STAR-RIS degrees of freedom for massive connectivity, but it also strengthens interference coupling between sensing and communication. As summarized in Table~\ref{tab:IV_E_summary_all}, RSMA designs mitigate multi-user interference via rate splitting while preserving sensing performance, providing flexible operating points in multi-user/multi-target settings. Fairness-aware NOMA further balances communication SINR and sensing quality (SCNR/CRLB), offering a systematic view of JCAS trade-offs, yet remaining sensitive to SIC ordering, CSI accuracy, and STAR-RIS coefficient-coupling constraints. A second trend is toward scalable multi-user designs via scheduling, clustering, and reduced-complexity updates: scheduling addresses time-varying user/target sets, while CB-NOMA with element-wise optimization reduces passive-beamforming complexity for large STAR-RIS sizes. Learning-assisted methods (DRL/meta-DRL) are increasingly used for mixed discrete--continuous decisions (e.g., element selection, pairing, and resource assignment) and cross-environment adaptation. Nevertheless, Table~\ref{tab:IV_E_summary_all} highlights persistent challenges, including robustness to imperfect/outdated CSI, SIC error propagation, inter-cluster interference control, and real-time feasibility under short coherence times. Future work should prioritize robust, low-overhead designs and experimentally grounded validation for large-scale multi-user STAR-RIS JCAS.

\subsection{STAR-RIS Enabled JCAS in High Mobility Scenarios and UAV Communications}

The integration of STAR-RIS into high-mobility platforms (e.g., vehicles and UAVs) is emerging as a key enabler for JCAS in dynamic environments, where fast time variations, blockage, and rapidly changing geometry challenge conventional sensing and communication designs. Recent works span terrestrial vehicular/millimeter-wave (mmWave) settings and aerial UAV-assisted JCAS, with emphasis on mobility-aware channel acquisition, beam tracking, and trajectory/resource adaptation.

\subsubsection{Vehicular and mmWave high-mobility frameworks}
In the vehicular context, \cite{Li_HM_GC,Li_HM_TC} investigated vehicle-mounted STAR-RIS-enabled JCAS designs for high-mobility mmWave operation. Their core idea is to exploit structured precoding/combining to extract cascaded channel parameters (e.g., delay/angle) and enable joint communication and localization, with STAR-RIS coefficients updated based on sensed mobility information to balance the localization--communication objectives. Although the framework demonstrates improved tracking and reliability compared to conventional RIS-based baselines, its performance depends on robust sparse recovery and channel identifiability under practical NLoS and rapidly varying conditions.

\subsubsection{Mobility-aware sensing of dynamic scatterers and beam tracking}
To address broader mobility-driven scenarios, \cite{MUYE-TVT} proposed an active STAR-RIS-aided JCAS framework that simultaneously serves indoor users while tracking dynamic outdoor scatterers. A key contribution is the integration of sensing into the STAR-RIS operation to filter relevant targets and enable beam prediction/tracking for improved communication reliability. The design couples mobility modeling (scatterer dynamics) with BS precoding and STAR-RIS phase updates, reducing training overhead via mismatch detection and collision prediction. This approach highlights the value of STAR-RIS-assisted sensing for proactive beam management, though its robustness may be sensitive to modeling assumptions (e.g., state transitions and RCS-based classification) under highly non-stationary mobility.

\subsubsection{Active STAR-RIS (ASTARS) for mobility-robust radar-centric sensing}
Complementing passive STAR-RIS designs, \cite{ge2024ofdm} studied an ASTARS-aided JCAS architecture in which communication users and moving sensing targets are placed on opposite sides of the surface to reduce mutual interference. Using OFDM signaling to combat frequency selectivity, the work maximizes radar SNR subject to communication SINR constraints via joint optimization of transmit beamforming and ASTARS amplitude/phase coefficients (solved numerically). The results show improved radar performance (range/velocity estimation) and BER trends under mobility, while also indicating that computational overhead can increase as the waveform and optimization complexity grow.
\begin{table*}[!t]
\centering
\caption{Summary of STAR-RIS enabled JCAS works in high-mobility and UAV-assisted scenarios.}
\label{tab:IV_C_summary}
\renewcommand{\arraystretch}{1.08}
\setlength{\tabcolsep}{5.0pt}


\begin{tabular}{|p{0.7cm}|p{2.7cm}|p{2cm}|p{2.3cm}|p{3.0cm}|p{1.9cm}|p{3cm}|}
\hline
\rowcolor{HeaderRed}
\textbf{Ref.} & \textbf{Mobility Scenario} & \textbf{Architecture / Protocol} & \textbf{Objective} & \textbf{Optimized Variables} & \textbf{Method} & \textbf{Notes / Metrics} \\
\hline

\rowcolor{BodyBlue}
\cite{Li_HM_GC} 2023 &
Vehicle-mounted mmWave, high mobility &
Vehicle-mounted STAR-RIS JCAS &
Joint localization, communication &
Precoding/combining, STAR-RIS phase updates &
Sparse recovery (e.g., OMP/MOMP), adaptive updates &
Tracking accuracy, reliability; sensitive to NLoS and identifiability \\
\hline

\rowcolor{BodyBlue}
\cite{Li_HM_TC} 2024 &
Vehicle-mounted mmWave, high mobility &
Vehicle-mounted STAR-RIS JCAS &
Joint localization, communication &
Precoding/combining, STAR-RIS phase updates &
Sparse recovery (e.g., OMP/MOMP), adaptive updates &
Tracking accuracy, reliability; sensitive to NLoS and identifiability \\
\hline

\rowcolor{BodyBlue}
\cite{MUYE-TVT} 2025 &
High-mobility mmWave with dynamic scatterers &
Active STAR-RIS JCAS with target filtering &
Beam prediction/tracking, comm enhancement &
BS precoding, STAR-RIS phase shifts (mobility-aware) &
State modeling, tracking filter, beam prediction &
Comm performance, tracking; depends on mobility/RCS modeling \\
\hline

\rowcolor{BodyBlue}
\cite{ge2024ofdm} 2024 &
High mobility (moving targets) &
ASTARS-aided JCAS (radar-centric), OFDM &
Max radar SNR with comm SINR constraints &
BS beamforming, ASTARS amplitude/phase &
Numerical optimization (CVX-type) &
Radar SNR, BER, range/velocity; higher computational overhead \\
\hline

\rowcolor{BodyBlue}
\cite{Zheng-TWC} 2024 &
Mobile users, bi-directional sensing &
Sensor-embedded STARS (TS protocol) &
Min CRB subject to QoS &
Waveform/power, STARS coefficients, sensor usage &
AO, SDP-based updates &
CRB/DoA, QoS; assumes ideal sync and low switching overhead \\
\hline

\rowcolor{BodyBlue}
\cite{Chen-CL} 2025 &
UAV DFRC JCAS (trajectory) &
STAR-RIS-assisted UAV DFRC (SRaDUI) &
Improve sensing efficiency and achievable rate &
UAV trajectory/hovering, scheduling, covariance, STAR-RIS coeffs &
AO, SCA/SDR decomposition &
Rate, sensing efficiency; depends on UAV mobility/energy assumptions \\
\hline

\rowcolor{BodyBlue}
\cite{HuangWCSP} 2024 &
UAV air-ground, near-field effects &
Semi-passive STAR-RIS, near-field JCAS &
Max WSR / balance sensing--comm &
Beamforming, STAR-RIS T/R matrices, UAV hovering location &
BCD, SDR/SCA &
WSR, sensing metric; highlights near-field modeling impact \\
\hline

\rowcolor{BodyBlue}
\cite{Yasoub-TVT} 2025 &
UAV dynamic environments (real-time adaptation) &
Learning-assisted STAR-RIS UAV JCAS &
Admission/control and JCAS trade-off under mobility &
Trajectory/resource actions, STAR-RIS configuration &
RDPG / DRL (meta-learning extension) &
Admission rate, trade-off; training overhead and robustness issues \\
\hline

\rowcolor{BodyBlue}
\cite{11148878} 2025 &
UAV dynamic environments (real-time adaptation) &
Learning-assisted STAR-RIS UAV JCAS &
Admission/control and JCAS trade-off under mobility &
Trajectory/resource actions, STAR-RIS configuration &
RDPG / DRL (meta-learning extension) &
Admission rate, trade-off; training overhead and robustness issues \\
\hline

\rowcolor{BodyBlue}
\cite{11174571} 2025 &
UAV dynamic environments (real-time adaptation) &
Learning-assisted STAR-RIS UAV JCAS &
Admission/control and JCAS trade-off under mobility &
UAV control actions + STAR-RIS configuration (and possibly scheduling) &
DRL (e.g., RDPG/TD3; meta-learning in some designs) &
Admission rate / sensing--comm trade-off; training overhead and robustness remain open \\
\hline

\end{tabular}
\end{table*}

\subsubsection{Bi-directional sensing and sensor-embedded STARS under mobility}
In \cite{Zheng-TWC}, a bi-directional sensing STARS architecture was proposed to overcome the unidirectional limitation of conventional RIS/STARS in mobile settings. By embedding micro-sensors and employing a time-switching protocol between transmission and reflection modes, the design supports full-space sensing/communication. CRB-driven optimization is performed to minimize the sensing error subject to QoS constraints using updates based on AO and SDP over the waveform/power and STARS coefficients. The work provides analytical insights (e.g., bounds on deployable sensors) but assumes ideal synchronization and negligible switching overhead, which should be carefully considered in practical implementations.

\subsubsection{UAV-assisted STAR-RIS JCAS via trajectory optimization and scheduling}
Moving to aerial mobility, \cite{Chen-CL} proposed a STAR-RIS-assisted DFRC UAV JCAS framework where the UAV performs ground target detection and multicast communication. The joint design optimizes UAV trajectory/hovering, target scheduling, transmit covariance, and STAR-RIS configuration using AO with SCA/SDR sub-steps, demonstrating improved sensing efficiency and achievable rates compared to conventional baselines. Similarly, \cite{HuangWCSP} considered STAR-RIS-enabled air-ground JCAS while explicitly accounting for near-field effects in large-scale deployments. The authors jointly optimized active beamforming, STAR-RIS transmission/reflection matrices, and hovering UAV locations via BCD with SDR/SCA, showing gains in both sensing and communication under near-field modeling.

\subsubsection{Learning-assisted UAV STAR-RIS JCAS under dynamic environments}
To enable real-time adaptation under rapidly changing mobility, \cite{Yasoub-TVT} formulated a STAR-RIS-aided UAV JCAS admission/control problem and solved it using a recurrent deep reinforcement learning approach, with a meta-learning extension for faster adaptation across mobility patterns. Complementing this direction, \cite{11174571} introduced a DRL-driven STAR-RIS-assisted UAV framework that integrates sensing, communication, and computation, where the learning agent adapts the actions of the system in response to time-varying conditions and coupled objectives. Similarly, \cite{11148878} investigated an aerial hybrid STAR-RIS-enabled JCAS architecture and adopted deep reinforcement learning to jointly handle the JCAS trade-off and STAR-RIS configuration under UAV mobility. Collectively, these learning-based designs highlight the promise of STAR-RIS for mobility-aware JCAS, while also motivating further study on training overhead, stability, robustness to imperfect CSI, and the feasibility of deploying iterative learning/optimization within short coherence intervals. 

\noindent\textbf{Discussion and insights:}
The studies in this subsection indicate that mobility fundamentally reshapes STAR-RIS-aided JCAS by tightly coupling sensing, channel acquisition, and resource control with rapidly varying geometry and Doppler. Terrestrial high-mobility works focus on estimating/tracking cascaded channel parameters (e.g., delay/angle) and leveraging sensing feedback to proactively update STAR-RIS configurations and beam directions, sustaining reliability under mmWave blockage and fast time variation. Active STAR-RIS/ASTARS designs can further enhance robustness by filtering/emphasizing relevant scatterers/targets, but their performance depends on motion/target models and can degrade under model mismatch. Sensor-embedded STARS architectures offer CRB-driven stabilization of sensing under mobility, yet synchronization and switching overhead may become key practical bottlenecks. In UAV-assisted scenarios, the dominant trend is joint trajectory/hovering and STAR-RIS configuration optimization. AO frameworks with SCA/SDR-type decompositions achieve strong performance but can be computationally heavy for real-time operation. Learning-based methods are therefore increasingly explored for fast adaptation in dynamic environments, particularly for mixed discrete--continuous decisions (e.g., admission control, scheduling, and coefficient updates). Overall, Table~\ref{tab:IV_C_summary} highlights open challenges in (i) robustness to imperfect/outdated CSI and model mismatch, (ii) complexity and update latency relative to channel coherence time, and (iii) scalable designs for larger user/target sets and large-scale STAR-RIS deployments in near-field and high-frequency regimes.





\section{Lessons Learned, Open Challenges and Research Directions}

This section consolidates the main takeaways from the surveyed literature and outlines the remaining obstacles and promising research directions for STAR-RIS-enabled JCAS. Although existing results demonstrate that STAR-RIS can substantially expand the feasible sensing--communication operating region through full-space wave manipulation, they also reveal that practical deployments require solving several remaining challenges. The following discussion synthesizes these insights and provides a forward-looking roadmap for future work. 

\subsection{Lessons Learned}
The reviewed studies collectively indicate that STAR-RIS introduces unique degrees of freedom that can be exploited to jointly improve communication reliability and sensing fidelity. At the same time, the literature suggests that many reported gains rely on idealized assumptions (e.g., perfect CSI and negligible control overhead), highlighting the need for more deployment-grounded designs.
 \begin{itemize}
    \item \emph{Full-space controllability is the core differentiator:} 
    The ability of STAR-RIS to simultaneously transmit and reflect incident signals enables true $360^\circ$ spatial coverage, overcoming the inherent half-space limitation of conventional reflective or transmissive RIS. This full-space controllability significantly improves service continuity for communication users and improves sensing visibility for targets located on either side of the surface. Moreover, it allows a single STAR-RIS to support heterogeneous users and sensing tasks in different spatial regions, reducing deployment density while increasing system flexibility and coverage robustness.

    \item \emph{Protocol choice governs the complexity--performance trade-off:} 
    The selection among ES, MS, and TS protocols plays a critical role in determining both achievable performance and implementation complexity. ES provides the highest design flexibility by enabling simultaneous control of transmitted and reflected signals at each element, but it introduces strongly coupled and highly non-convex optimization problems. In contrast, MS and TS simplify hardware control and optimization by decoupling transmission and reflection in space or time, respectively, at the expense of reduced array gain, additional scheduling constraints, or strict synchronization requirements.

    \item \emph{Joint design is essential rather than optional:} 
    The surveyed literature consistently demonstrates that independent or sequential optimization of active beamforming and STAR-RIS coefficients leads to suboptimal performance. In contrast, end-to-end joint optimization significantly improves both communication reliability and sensing accuracy, particularly when explicit sensing constraints, such as minimum SINR, Cramér–Rao bounds (CRB), or signal-to-clutter-and-noise ratio (SCNR), are incorporated alongside communication quality-of-service (QoS) requirements. This highlights the intrinsically coupled nature of sensing and communication in STAR-RIS-enabled JCAS systems.

    \item \emph{Waveform shaping complements passive control:} 
Though STAR-RIS offers considerable passive control over the propagation environment, waveform design is an essential complementary degree of freedom. Constructive interference, ambiguity in sensing, and robustness to clutter can be improved substantially by the explicit design of JCAS, constant-modulus, pulsed, or space-time coded signal waveforms, such as dual-functional radar-communication (DFRC) integrated waveforms or JCAS waveforms. This is especially the case for interference and clutter in the presence of multiple users and multiple targets, as the waveform design structure can interact with and determine multiple user interference and the resolution of the sensing.

    \item \emph{Scalability is a limiting factor in algorithmic design:} 
  Most current STAR-RIS-enabled JCAS solutions heavily depend on computation-heavy techniques like semidefinite relaxation (SDR), successive convex approximation (SCA), or alternating optimization (AO). While effective for moderate system sizes, these methods do not scale well with the number of RIS elements, users, or sensing targets, and do not easily accommodate rapid channel changes. This highlights the importance of lower complexity algorithm architectures, layered optimization, and real-time feasible learning-based control.

    \item \emph{Mobility and near-field effects change the design regime:} 
    In vehicular and UAV-assisted networks and in scenarios with large aperture STAR-RIS operating in the electromagnetic near-field, the standard far-field and quasi-static channel assumptions become overly simplified. Channel non-stationarity, spherical wavefront propagation, and spatially Varying array responses impact the sensing models and the optimal configurations of the STAR-RIS. All these effects call for the design of mobility-aware systems, predictive control mechanisms, and the use of models for signal and sensing that accounts the near-field in order to fully harness the potential of STAR-RIS-enabled JCAS.
\end{itemize}


\subsection{Open Challenges}
Even though we have come a long way, some important problems still have to be solved, which hinder turning theories into practical applications. These problems include the practicality of hardware, problems related to channel gain, system-level overhead, and the robustness needed in rapidly changing environments.
 \begin{itemize}
    \item \emph{Accurate and scalable CSI acquisition:} 
    Every JCAS system that employs STAR-RIS includes a series of interconnected components including: the transmitter, STAR-RIS, STAR-RIS—receiver, and the STAR-RIS—target links, and this applies to both transmitted and reflected signal pathways. Obtaining precise CSI for these links is particularly arduous considering user mobility, dynamic blockage, and the limited availability of pilot resources. The problem worsens when we consider large numbers of STAR-RIS elements, N, and when the system is further extended to include multiple communicative users and sensing targets. Additionally, there is a close interdependence between sensing parameters (e.g., target location or Doppler) and communication-related CSI, which necessitates joint estimation techniques while remaining within the bounds of realistic computational limits.

    \item \emph{Coefficient coupling and hardware constraints:} 
    The majority of theoretical studies outline uncorrelated and uninterrupted control of the amplitude and phase for the reflected and transmitted coefficients. However, there are several practical considerations STAR-RIS elements must take into account, such as coupled amplitude-phase response, discrete phase quantization, inter-element mutual coupling, and insertion and leakage losses. These factors can impede their performance. These hardware constraints lead to the inability to perform the desired wave manipulation and greatly reduce the potential gains in sensing and communication. To be able to actually deploy the proposed systems, there is a need to create models and compensate for the hardware constraints.

    \item \emph{Control signaling and latency overhead:} 
   The effective operation of STAR-RIS requires updates of element configurations due to changes in channels, sensing goals, and network status. For wide surfaces with precise control over individual elements, this could lead to a considerable control-plane signaling overhead. With such surfaces, the TS protocol introduces additional extreme time synchronization constraints between the transmission and reflection phases. When the sensing and communication functions necessitate quick responses, as in the case with vehicles or UAVs, the latency and signaling overhead may counterbalance the expected performance improvements of STAR-RIS-supported JCAS.

    \item \emph{Robustness under environmental dynamics:} 
    Non-stationary scatterers, intermittent line-of-sight conditions, and time-varying clutter are some of the major characteristics of real-world environments. JCAS systems are particularly STAR-RIS enabled. This is due to the fact that both sensing and communication are dependent on controlling the waves in a precise manner. When there are changes to the environment that utilized the predicted model of a channel or target, performance rapidly degrades. There have been a number of robust optimization frameworks developed that focus on statistical or bounded uncertainty, but these tend to have high computational complexity and are still difficult to apply to large-scale systems in real time.

    \item \emph{Joint interference management for sensing and communication:} 
    The JCAS systems are interfered in several ways: multi-user communication interference, sensing echo interference, and clutter leakage. These interference components are closely related, especially under shared waveforms and STAR-RIS-assisted propagation. Framing the design of either suppressive or exploitative interference strategies for joint performance gains remains an open challenge, especially for dense deployments with a large number of users, targets, and reflective objects. The problem of interference coupling within the sensing and communication domains remains underexplored.

    \item \emph{Energy and thermal constraints for active STAR-RIS:} 
    Active STAR-RIS architectures add amplifiers or active circuits to mitigate path loss and improve coverage. This approach, while providing substantial performance gains, comes at the expense of added noise, increased power consumption, heat dissipation, and possible stability complications. Therefore, energy efficiency and thermal management become increasingly important design constraints, particularly for large areas or when the system must operate continuously. The system-level models that jointly consider the benefits of amplification versus the drawbacks of noise, along with power and thermal constraints, are still developing.

    \item \emph{Evaluation methodology and reproducibility:} 
    The existing body of work does not offer a comprehensive assessment of STAR-RIS-enabled JCAS, and it has yet to reach a distinct consensus on what it terms representative benchmark scenarios, balanced quantification of sensing metrics, channel and hardware impairment models, shared datasets, and so forth. Consequently, it is often challenging and at times even misleading to assess varying solutions across studies. Thus, it is critical to provide open assessment platforms and reproducible benchmarking criteria to evaluate advancements and surface solutions that are scalable and robust.

    \item \emph{Cross-layer integration and system-level complexity:} 
    While most current studies target physical-layer designs, the upper-layer considerations like scheduling, mobility management, and network coordination are either oversimplified or ignored. STAR-RIS-enabled JCAS, functioning in the field, must deal with complicated multi-layer network structures. Decisions made at the physical layer are intertwined with the medium access control, routing, and application layers. Therefore, handling and understanding such cross-layer interdependencies is fundamental to achieving end-to-end performance improvements.

    \item \emph{Security, privacy, and sensing integrity:} 
    Because of its ability to be programmed, STAR-RIS can be reconfigured, spoofed, and can eavesdrop via rogue reflections or transmissions. Additionally, sensing capabilities can create privacy risks from environmental and user surveillance. In STAR-RIS-enabled JCAS systems, maintaining secure control, protecting sensing, and keeping privacy in STAR-RIS systems is still mostly unresolved, especially in large and decentralized systems.

\end{itemize}


\subsection{Research Directions}
Addressing the above challenges requires advances that integrate electromagnetic design, signal processing, and network control. The following research directions represent high-impact opportunities to mature STAR-RIS-enabled JCAS toward practical 6G systems.
 \begin{itemize}
    \item \emph{Overhead-aware and task-driven channel estimation:} 
    Future research should move beyond exhaustive channel estimation toward overhead-aware and task-driven strategies tailored to JCAS objectives. By exploiting sensing information, spatial sparsity, and geometric structure, sensing-assisted and compressed CSI acquisition schemes can significantly reduce pilot overhead. Rather than estimating full cascaded channels, task-oriented training should focus on extracting only the dominant parameters relevant to communication and sensing performance, such as key propagation paths, angles, delays, or target features. This paradigm is particularly important for large STAR-RIS deployments and highly dynamic environments.

    \item \emph{Hardware-consistent STAR-RIS modeling and design:} 
    Bridging the gap between theory and practice requires STAR-RIS models that faithfully capture hardware realities, including amplitude–phase coupling, finite-resolution control, insertion loss, mutual coupling, and calibration errors. Optimization frameworks must explicitly account for these constraints to avoid overly optimistic performance predictions. In parallel, new metasurface architectures should be explored to enable more independent and predictable transmission/reflection control while strictly respecting power conservation laws. Such hardware-aware design is essential for scalable and reliable deployment.

    \item \emph{Low-complexity real-time optimization:} 
    To support fast-varying channels and time-critical sensing tasks, future STAR-RIS-enabled JCAS systems require optimization methods with low computational complexity and predictable runtime. Promising directions include closed-form or semi-closed-form updates, first-order and incremental optimization techniques, and hybrid learning–optimization approaches that leverage offline training with online lightweight adaptation. Ensuring bounded convergence time and robustness is critical for real-time operation in large-scale systems.

    \item \emph{Integrated waveform--beam--surface co-design:} 
    A holistic design that jointly optimizes waveform structure, active beamforming, and STAR-RIS coefficients is essential for fully exploiting the available degrees of freedom. Unified performance metrics, such as communication rate, CRB, or squared position error bound (SPEB), and signal-to-clutter-and-noise ratio (SCNR), should guide this co-design. Special attention should be given to mobility-resilient waveforms, including pulsed, space–time coded, and OTFS-like designs, which can enhance both sensing robustness and communication reliability.

    \item \emph{Mobility-aware predictive control:} 
    In vehicular and UAV-assisted scenarios, reactive reconfiguration of STAR-RIS may be insufficient due to rapid channel variations. Predictive control frameworks that combine tracking, environmental awareness, and mobility prediction can proactively configure beams and surface coefficients. Techniques such as Bayesian filtering, map-assisted modeling, and learning-based forecasting can reduce retraining frequency, lower signaling overhead, and improve reliability under high mobility and intermittent connectivity.

    \item \emph{Near-field and XL-array JCAS with STAR-RIS:} 
    As STAR-RIS apertures grow large and operate at higher frequencies, near-field effects and spatial non-stationarity become dominant. Future work should extend JCAS models to spherical-wave propagation and develop geometry-aware sensing and communication formulations. This includes joint design of sensor placement on or near the STAR-RIS, near-field beam focusing, and location-aware estimation techniques that exploit the unique spatial resolution offered by extremely large (XL) surfaces.

    \item \emph{Security, privacy, and sensing integrity:} 
    STAR-RIS-enabled JCAS introduces new opportunities and risks at the physical layer. Future research should investigate secure waveform and surface configurations that enhance confidentiality while preserving sensing performance. Privacy-preserving sensing mechanisms are also needed to prevent unauthorized inference of sensitive environmental or user information. Additionally, robustness against adversarial attacks, such as spoofing, jamming, and malicious RIS reconfiguration, must be systematically addressed.

    \item \emph{System-level and cross-layer design:} 
    Practical deployment of STAR-RIS-enabled JCAS requires system-level studies that go beyond isolated links. Future research should consider multi-cell and multi-RIS networks, accounting for scheduling, coordination, and backhaul constraints. Architectures that balance centralized optimization with distributed and autonomous control are particularly attractive, enabling scalability while limiting control overhead and latency.

    \item \emph{Prototyping and standardized benchmarking:} 
    To accelerate technology maturation, experimental prototyping and standardized benchmarking are indispensable. Building hardware testbeds and real-world demonstrators will help validate theoretical gains under practical impairments. In parallel, the community should establish reproducible benchmarks, including reference channel models, impairment profiles, sensing metrics, and evaluation protocols, to enable fair comparison and foster cumulative progress.
\end{itemize}
To conclude, STAR-RIS-enabled JCAS offers a compelling pathway toward perceptive and reconfigurable 6G networks. However, realizing its full potential hinges on advancing hardware-realistic modeling, scalable real-time control, and rigorous experimental validation across representative deployment scenarios.

\section{Conclusion}
This paper focused on the STAR-RIS-enabled JCAS systems and their pivotal role as the breakthrough technology for 6G wireless networks. A STAR-RIS and JCAS system was primarily chosen due to the basic principles associated with them, including the various models, protocols, hardware designs, and metrics related to performance. Efficient solutions were also systemically categorized for various models. From the recent studies associated with waveform design, beamforming, resource allocation, and optimization with machine learning, we pointed out the unique ability of STAR-RIS to enhance communication by adding dimensions to the sensing through controlling the entire space. This capability is as a result of the design of STAR-RIS, therefore simplifying channel acquisition. We also provided a summarized version of the lessons learned from the systems and the various challenges associated with channel acquisition, the limitations of the hardware, and a few others. The challenges provided a direction of where the research should be focused. This paper primarily focuses on providing a solid reference from the survey to the researchers and practitioners the pathways for the creation of the flexible, efficient, and the intelligent STAR-RIS - JCAS systems for the wireless networks of the future.

\ifCLASSOPTIONcaptionsoff
  \newpage
\fi

{
\def\baselinestretch{0.9}
\bibliographystyle{IEEEtran}
\bibliography{main}
}

\end{document}